# Asymptotic confidence intervals for the difference and the ratio of the weighted kappa coefficients of two diagnostic tests subject to a paired design

José Antonio Roldán-Nofuentes[a], Saad Bouh Sidaty-Regad[b]

[a]Biostatistics, School of Medicine, University of Granada, Spain

[b]Public Health and Epidemiology, School of Medicine, University of Nouakchott, Mauritania

Email: jaroldan@ugr.es, sidaty_saad@yahoo.com

**Abstract.** The weighted kappa coefficient of a binary diagnostic test is a measure of the beyond-chance agreement between the diagnostic test and the gold standard, and depends on the sensitivity and specificity of the diagnostic test, on the disease prevalence and on the relative importance between the false positives and the false negatives. This article studies the comparison of the weighted kappa coefficients of two binary diagnostic tests subject to a paired design through confidence intervals. Three asymptotic confidence intervals are studied for the difference between the parameters and five other intervals for the ratio. Simulation experiments were carried out to study the coverage probabilities and the average lengths of the intervals, giving some general rules for application. A method is also proposed to calculate the sample size necessary to compare the two weighted kappa coefficients through a confidence interval. A program in R has been written to solve the problem studied and it is available as supplementary material. The results were applied to a real example of the diagnosis of malaria.

**Keywords:** weighted kappa coefficient, paired design, binary diagnostic test.

**Mathematics Subject Classification**: 62P10, 6207.

# 1. Introduction

A diagnostic test is medical test that is applied to an individual in order to determine the presence or absence of a disease. When the result of a diagnostic test is positive (indicating the presence of the disease) or negative (indicating its absence), the diagnostic test is called a binary diagnostic test (*BDT*) and its accuracy is measured in terms of two fundamental parameters: sensitivity and specificity. Sensitivity (*Se*) is the probability of the *BDT* result being positive when the individual has the disease, and specificity (*Sp*) is the probability of the *BDT* result being negative when the individual does not have the disease. Sensitivity is also called true positive fraction (*TPF*) and specificity is also called true negative fraction (*TNF*), verifying that $TPF = 1 - FNF$ and that $TNF = 1 - FPF$, where *FNF* (*FPF*) is the false negative (positive) fraction. The accuracy of a *BDT* is assessed in relation to a gold standard (*GS*), which is a medical test that objectively determines whether or not an individual has the disease. When considering the losses of an erroneous classification with the *BDT*, the performance of the *BDT* is measured in terms of the weighted kappa coefficient (Kraemer et al, 1990; Kraemer, 1992; Kraemer et al, 2002). The weighted kappa coefficient depends on the *Se* and *Sp* of the *BDT*, on the disease prevalence (*p*) and on the relative importance between the false positives and the false negatives (weighting index *c*). The weighted kappa coefficient is a measure of the beyond-chance agreement between the *BDT* and the *GS*.

Furthermore, the comparison of the performance of two *BDTs* is an important topic in the study of Statistical Methods for Diagnosis in Medicine. The comparison of two *BDTs* can be made subject to two types of sample designs: unpaired design and paired design. In the book by Pepe (2003) we can see a broad discussion about both types of sample designs. Summing up, subject to an unpaired design each individual is tested with a single *BDT*, whereas subject to a paired design each individual is tested with the two *BDT*s. Consequently, unpaired design consists of applying a *BDT* to a sample of $n_1$ individuals and the other *BDT* to another sample of $n_2$ individuals; paired design consists of applying both *BDTs* to all of the individuals of a sample sized *n*. The comparative studies based on a paired design are more efficient from a statistical point of view than the studies based on an unpaired design, since it minimizes the impact of the between-individual variability. Therefore, in this article we focus on paired design. Subject to this type of design, Bloch (1997) has studied an asymptotic hypothesis test to compare the weighted kappa coefficients of two *BDTs*. Nevertheless, if the hypothesis test is significant, this method does not allow us to assess how much bigger one weighted kappa coefficient is compared to another one, and it is necessary to estimate this effect through confidence intervals (*CIs*). Thus, the objective of our study is to compare the weighted kappa coefficients of two *BDTs* through *CIs*. Frequentist and Bayesian *CIs* have been studied for the difference and for the ratio of the two weighted kappa coefficients. If a *CI* for the difference (ratio) does not contain the zero (one) value, then we reject the equality between the two weighted kappa coefficients and we estimate how much bigger one coefficient is than another one. Consequently, our study is an extension of the Bloch method to the situation of the *CIs*. We have also dealt with the problem of calculating the sample size to compare the two parameters through a *CI*.

The manuscript is structured in the following way. In Section 2, we explain the weighted kappa coefficient of a *BDT* and we relate the comparison of the weighted kappa coefficients of two *BDTs* with the relative true (false) positive fraction of the two *BDTs*. Section 3 summarizes the Bloch method and we propose *CIs* for the difference and the ratio of the weighted kappa coefficients of two *BDTs* subject to a paired design. In Section 4, simulation experiments are carried out to study the asymptotic behaviour of the proposed *CIs*, and some general rules of application are given. In Section 5, we propose a method to calculate the sample size necessary to compare the two weighted kappa coefficients through a *CI*. In Section 6, a programme written in *R* is presented to solve the problems posed in this manuscript. In Section 7, the results were applied to a real example on the diagnosis of malaria, and in Section 8 the results are discussed.

## 2. Weighted kappa coefficient

Let us consider a *BDT* that is assessed in relation to a *GS*. Let $L$ ($L'$) the loss which occurs when for a diseased (non-diseased) individual the *BDT* gives a negative (positive) result. Therefore, the loss $L$ ($L'$) is associated with a false negative (positive). If an individual (with or without the disease) is correctly diagnosed by the *BDT* then $L = L' = 0$. Let $D$ be the variable that models the result of the *GS*: $D = 1$ when an individual has the disease and $D = 0$ when this is not the case. Let $p = P(D = 1)$ be the prevalence of the disease and $q = 1 - p$. Let $T$ be the random variable that models the result of the *BDT*: $T = 1$ when the result of the *BDT* is positive and $T = 0$ when the result is negative. Table 1 shows the losses and the probabilities associated with the assessment of a *BDT* in relation to a *GS*, and the probabilities when the *BDT* and the *GS* are independent, i.e. when $P(T = i | D = j) = P(T = i)$. Multiplying each loss in the $2 \times 2$ table by its corresponding probability and adding up all the terms, we find $p(1 - Se)L + q(1 - Sp)L'$, a term that is defined as expected loss. Therefore, the expected loss is the loss that occurs when erroneously classifying with the *BDT* an individual with or without the disease. Moreover, if the *BDT* and the *GS* are independent, multiplying each loss by its corresponding probability (subject to the independence between the *BDT* and the GS) and adding up all of the terms we find $p[p \times (1-Se) + q \times Sp]L + q[p \times Se + q \times (1-Sp)]L'$, a term that is defined as random loss. Therefore, the random loss is the loss that occurs when the *BDT* and the *GS* are independent. The independence between the *BDT* and the *GS* is equivalent to the Youden index of the *BDT* being equal to zero i.e. $Se + Sp - 1 = 0$, and is also equivalent to the expected loss being equal to the random loss. In terms of expected and random losses, the weighted kappa coefficient of a *BDT* is defined as

$$\kappa = \frac{\text{Random loss} - \text{Expected loss}}{\text{Random loss}}.$$

Substituting in this equation each loss with its expression, the weighted kappa coefficient of a *BDT* is expressed (Kraemer et al, 1990; Kraemer, 1992; Kraemer et al, 2002) as

$$\kappa(c) = \frac{pqY}{p(1-Q)c + qQ(1-c)}, \qquad (1)$$

where $Y = Se + Sp - 1$ is the Youden index, $Q = pSe + q(1-Sp)$ is the probability that the *BDT* result is positive, and $c = L/(L' + L)$ is the weighting index. The weighting index $c$ is a measure of the relative importance between the false positives and the false negatives. For example, let us consider the diagnosis of breast cancer using as a diagnostic mammography test. If the mammography test is positive in a woman that does not have cancer (false positive), the woman will be given a biopsy that will give a negative result. The loss $L'$ is determined from the economic costs of the diagnosis and also from the risk, stress, anxiety, etc., caused to the woman. If the mammography test is negative in a woman who has breast cancer (false negative), the woman may be diagnosed at a later stage, but the cancer may spread, and the possibility of the treatment being successful will have diminished. The loss $L$ is determined from these considerations. The losses $L$ and $L'$ are measured in terms of economic costs and also from risks, stress, etc., which is why in practice their values cannot be determined. Therefore, as loss $L$ cannot be determined, $L$ is substituted by the importance that a false positive has for the clinician; in the same way, as loss $L'$ cannot be determined, then $L'$ is substituted by the importance that a false negative has for the clinician. The value of the weighting index $c$ will depend therefore on the relative importance between a false positive and a false negative. If the clinician has greater concerns about false positives, as it is the situation in which the $BDT$ is used as a definitive test prior to a treatment that involves a risk for the individual (e.g., a definitive test prior to a surgical operation), then $0 \leq c < 0.5$. If the clinician is more concerned about false negatives, as in a screening test, then $0.5 < c \leq 1$. The index $c$ is equal to 0.5 when the clinician considers that the false negatives and the false positives have the same importance, in which case $\kappa(0.5)$ is the Cohen kappa coefficient. Weighting index $c$ quantifies the relative importance between a false positive and a false negative, but it is not a measure that quantifies how much bigger the proportion of false positives is compared to the false negatives. If $c = 0$ then

$$\kappa(0) = \frac{Sp - (1-Q)}{Q} = \frac{p(1 - FNF - FPF)}{p(1 - FNF) + qFPF}, \qquad (2)$$

which is the chance corrected specificity according to the kappa model. If $c = 1$ then

$$\kappa(1) = \frac{Se - Q}{1 - Q} = \frac{q(1 - FNF - FPF)}{pFNF - q(1 - FPF)}, \qquad (3)$$

which is the chance corrected sensitivity according to the kappa model. A low (high) value of $\kappa(1)$ will indicate that the value of $FNF$ is high (low), and a low (high) value of $\kappa(0)$ will indicate that the value of $FPF$ is high (low). The weighted kappa coefficient can be written as

$$\kappa(c) = \frac{pc(1-Q)\kappa(1) + q(1-c)Q\kappa(0)}{pc(1-Q) + q(1-c)Q}, \quad (4)$$

which is a weighted average of $\kappa(0)$ and $\kappa(1)$. Therefore, the weighted kappa coefficient is a measure that considers the proportion of false negatives (*FNF*) and the proportion of false positives (*FPF*). Moreover, for a set value of the *c* index and of the accuracy (*Se* and *Sp*) of the *BDT*, the weighted kappa coefficient strongly depends on the disease prevalence among the population being studied, and its value increases when the disease prevalence increases. The weighted kappa coefficient is a measure of the beyond-chance agreement between the *BDT* and the *GS*. The properties of the kappa coefficient can be seen in the manuscript of Roldán-Nofuentes and Amro (2018).

Table 1. Losses and probabilities.

| | Losses (Probabilities) | | |
|---|---|---|---|
| | $T=1$ | $T=0$ | Total |
| $D=1$ | $0\ (p\times Se)$ | $L\ (p\times(1-Se))$ | $L\ (p)$ |
| $D=0$ | $L'\ (q\times(1-Sp))$ | $0\ (q\times Sp)$ | $L'\ (q)$ |
| Total | $L'\ (Q = p\times Se + q\times(1-Sp))$ | $L\ (1-Q = p\times(1-Se) + q\times Sp)$ | $L+L'\ (1)$ |
| | Probabilities when the *BDT* and the *GS* are independent | | |
| | $T=1$ | $T=0$ | Total |
| $D=1$ | $p\times Q$ | $p\times(1-Q)$ | $p$ |
| $D=0$ | $q\times Q$ | $q\times(1-Q)$ | $q$ |
| Total | $Q$ | $1-Q$ | $1$ |

The weighted kappa coefficient is a valid parameter to assess and compare the performance of *BDTs* (Kraemer et al, 1990; Kraemer, 1992; Kraemer et al, 2002; Bloch, 1997; Roldán-Nofuentes et al, 2009; Roldán-Nofuentes and Amro, 2018).

When comparing the accuracies of two *BDTs*, Pepe (2003) recommends using the parameters $rTPF_{12} = \frac{Se_1}{Se_2}$ and $rFPF_{12} = \frac{FPF_1}{FPF_2}$, where $FPF_h = 1 - Sp_h$, with $h=1,2$. If $rTPF_{12} > 1$ then the sensitivity of Test 1 is greater than that of Test 2, and if $rFPF_{12} > 1$ then the *FPF* of Test 1 is greater than that of Test 2 (the specificity of Test 2 is greater than that of Test 1). The comparison of the weighted kappa coefficients of two *BDTs* can be related to the previous measures, and these have an important effect on the comparison of $\kappa_1(c)$ and $\kappa_2(c)$. From now onwards, it is considered that $0 < Se_h < 1$, $0 < Sp_h < 1$ and $0 < p < 1$, with $h = 1, 2$. Let us consider the subindexes $i$ and $j$, in such a way that if $i = 1$ ($i = 2$) then $j = 2$ ($j = 1$). It is obvious that if $rTPF_{ij} = rFPF_{ij} = 1$ then $Se_1 = Se_2$ and $Sp_1 = Sp_2$, and that therefore $\kappa_1(c) = \kappa_2(c)$ with $0 \le c \le 1$. Let

$$c' = \frac{(1-p)\left[Se_2(1-Sp_1) - Se_1(1-Sp_2)\right]}{p(Se_1 - Se_2) + (1-Sp_1)(Se_2 - p) - (1-Sp_2)(Se_1 - p)}. \tag{5}$$

In terms of $rTPF_{ij}$ and $rFPF_{12}$ the following rules are verified to compare $\kappa_1(c)$ and $\kappa_2(c)$:

a) If $rTPF_{ij} \ge 1$ and $rFPF_{ij} < 1$, or $rTPF_{ij} > 1$ and $rFPF_{ij} \le 1$, then $\kappa_i(c) > \kappa_j(c)$ for $0 \le c \le 1$.

b) If $rTPF_{ij} > 1$ and $rFPF_{ij} > 1$, then:

  b.1) $\kappa_i(c) > \kappa_j(c)$ if $0 < c' < c \le 1$

  b.2) $\kappa_i(c) < \kappa_j(c)$ if $0 \le c < c' < 1$

  b.3) $\kappa_1(c) = \kappa_2(c)$ if $c = c'$, with $0 < c' < 1$

  b.4) $\kappa_i(c) > \kappa_j(c)$ for $0 \le c \le 1$ if $c' < 0$ (or $c' > 1$) and $rTPF_{ij} > rFPF_{ij} > 1$

  b.5) $\kappa_i(c) < \kappa_j(c)$ for $0 \le c \le 1$ if $c' < 0$ (or $c' > 1$) and $rFPF_{ij} > rTPF_{ij} > 1$

c) If $rTPF_{ij} < 1$ and $rFPF_{ij} < 1$, then:

  c.1) $\kappa_i(c) > \kappa_j(c)$ if $0 \le c < c' < 1$

  c.2) $\kappa_i(c) < \kappa_j(c)$ if $0 < c' < c \le 1$

  c.3) $\kappa_2(c) = \kappa_1(c)$ if $c = c'$, with $0 < c' < 1$

c.4) $\kappa_i(c) > \kappa_j(c)$ for $0 \leq c \leq 1$ if $c' < 0$ (or $c' > 1$) and $rTPF_{ij} > rFPF_{ij} > 1$

c.5) $\kappa_i(c) < \kappa_j(c)$ for $0 \leq c \leq 1$ if $c' < 0$ (or $c' > 1$) and $rFPF_{ij} > rTPF_{ij} > 1$

The demonstrations can be seen in the Appendix A of the supplementary material. Regarding $c'$, this is obtained solving the equation $\kappa_1(c) - \kappa_2(c) = 0$ in $c$. The graphs in Figure 1 show how $\kappa_1(c)$ (on a continuous line) and $\kappa_2(c)$ (on a dotted line) vary depending on the weighting index $c$, taking as prevalence $p = \{5\%, 25\%, 50\%, 75\%\}$, for $Se_1 = 0.80$, $Sp_1 = 0.95$, $Se_2 = 0.90$ and $Sp_2 = 0.85$. These graphs correspond to the case in which $rTPF_{12} < 1$ and $rFPF_{12} < 1$, and therefore $\kappa_1(c) > \kappa_2(c)$ when $c < c'$, and $\kappa_2(c) > \kappa_1(c)$ when $c > c'$, and $c'$ is equal to 0.95 when $p = 5\%$, 0.75 when $p = 25\%$, 0.50 when $p = 50\%$ and 0.25 when $p = 75\%$. If the clinician considers that a false positive is 1.5 times more important than a false negative, then $c = 0.4$ and $\kappa_1(c) > \kappa_2(c)$ in the population with $p = \{5\%, 25\%, 50\%\}$ and $\kappa_2(c) > \kappa_1(c)$ in the population with $p = 75\%$. If in the population with $p = 75\%$ the clinician has a greater concern about a false positive than a false negative $(0 \leq c < 0.5)$, then $\kappa_1(c) > \kappa_2(c)$ if $0 \leq c < 0.25$ and $\kappa_2(c) > \kappa_1(c)$ if $0.25 < c < 0.5$; in the populations with $p = \{5\%, 25\%, 50\%\}$, $\kappa_1(c) > \kappa_2(c)$ when $0 \leq c < 0.5$.

We will now study the comparison of the weighted kappa coefficients of two *BDTs* through *CIs* subject to a paired design.

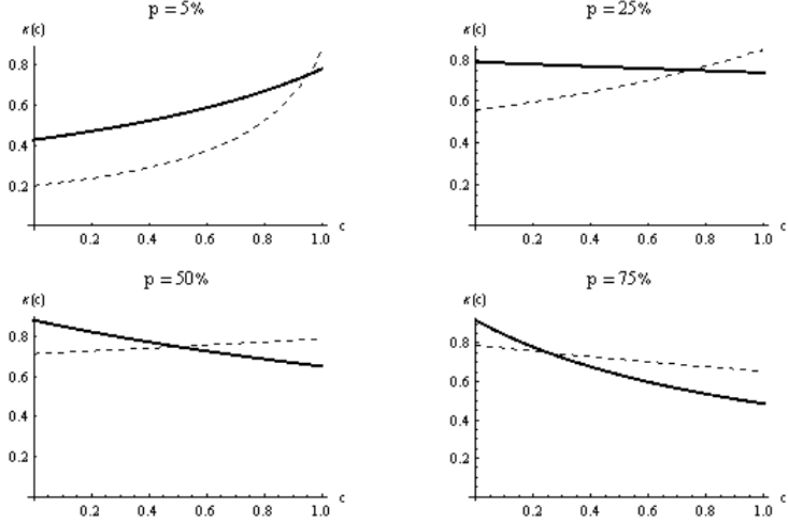

Figure 1. Weighted kappa coefficients with $rTPF_{12} < 1$ and $rFPF_{12} < 1$.
$Se_1 = 0.80 \quad Sp_1 = 0.95 \quad Se_2 = 0.90 \quad Sp_2 = 0.85 \quad rTPF_{12} = 0.89 \quad rFPF_{12} = 0.33$

## 3. Confidence intervals

Let us consider two *BDTs* which are assessed in relation to the same *GS*. Let $T_1$ and $T_2$ be the random binary variables that model the results of each *BDT* respectively. Let $Se_h$ and $Sp_h$ be the sensitivity and specificity of the *h*th *BDT*, with $h = 1, 2$. Table 2 (Observed frequencies) shows the frequencies that are obtained when both *BDTs* and the *GS* are applied to all the individuals in a random sample sized *n*. The frequencies $s_{ij}$ and $r_{ij}$ are the product of a multinomial distribution whose probabilities are also shown in Table 1 (Theoretical probabilities), where $p_{ij} = P(D = 1, T_1 = i, T_2 = j)$ and $q_{ij} = P(D = 0, T_1 = i, T_2 = j)$, with $i, j = 0, 1$. The probability of the two *BDTs* being positive when an individual has the disease is $Se_1 Se_2 + \varepsilon_1$, where $\varepsilon_1$ is the covariance or dependence factor between the two *BDTs* when $D = 1$; and the probability of the two *BDTs* being negative when an individual does not have the disease is $(1 - Sp_1)(1 - Sp_2) + \varepsilon_0$, where $\varepsilon_0$ is the covariance or dependence factor between the two *BDTs* when $D=0$. This model is known as the Vacek (1985) conditional dependence model. Applying this model, the probabilities $p_{ij}$ and $q_{ij}$ are written as

$$p_{ij} = p\left[ Se_1^i (1-Se_1)^{1-i} Se_2^j (1-Se_2)^{1-j} + \delta_{ij}\varepsilon_1 \right] \tag{6}$$

and

$$q_{ij} = q\left[ Sp_1^{1-i} (1-Sp_1)^i Sp_2^{1-j} (1-Sp_2)^j + \delta_{ij}\varepsilon_0 \right], \tag{7}$$

where $\delta_{ij}=1$ if $i=j$ and $\delta_{ij}=-1$ if $i \neq j$, with $i,j=0,1$. It is verified that $0 \leq \varepsilon_1 \leq Min\{Se_1(1-Se_2), Se_2(1-Se_1)\}$ and $0 \leq \varepsilon_0 \leq Min\{Sp_1(1-Sp_2), Sp_2(1-Sp_1)\}$. If $\varepsilon_1 = \varepsilon_0 = 0$ then the two *BDTs* are conditionally independent on the disease. In practice, the assumption of conditional independence is not realistic, and so $\varepsilon_1 > 0$ and/or $\varepsilon_0 > 0$.

Let $\boldsymbol{\pi} = (p_{11}, p_{10}, p_{01}, p_{00}, q_{11}, q_{10}, q_{01}, q_{00})^T$ be the vector of probabilities of the multinomial distribution, and it is verified that $p = \sum_{i,j=0}^{1} p_{ij}$ and $q = 1 - p = \sum_{i,j=0}^{1} q_{ij}$. The maximum likelihood estimators of the probabilities are $\hat{p}_{ij} = s_{ij}/n$ and $\hat{q}_{ij} = r_{ij}/n$.

The rules given in Section 2 about the effect of *rTPF* and *rFPF* on the comparison of $\kappa_1(c)$ and $\kappa_2(c)$ are theoretical rules that can be applied to the estimators, but they cannot guarantee that one weighted kappa coefficient will be higher than another. This question should be studied through hypothesis tests and confidence intervals. The Bloch method to compare the weighted kappa coefficients of two *BDTs* subject to a paired design is summarized below, and different *CIs* are proposed to compare these parameters subject to the same type of sample design.

Table 2. Observed frequencies and theoretical probabilities when two *BDTs* are compared in relation to a *GS* subject to a paired design.

| | Observed frequencies (Theoretical probabilities) | | | | |
|---|---|---|---|---|---|
| | $T_1 = 1$ | | $T_1 = 0$ | | |
| | $T_2 = 1$ | $T_2 = 0$ | $T_2 = 1$ | $T_2 = 0$ | Total |
| $D = 1$ | $s_{11}$ $(p_{11})$ | $s_{10}$ $(p_{10})$ | $s_{01}$ $(p_{01})$ | $s_{00}$ $(p_{00})$ | $s$ $(p)$ |
| $D = 0$ | $r_{11}$ $(q_{11})$ | $r_{10}$ $(q_{10})$ | $r_{01}$ $(q_{01})$ | $r_{00}$ $(q_{00})$ | $r$ $(q)$ |
| Total | $s_{11} + r_{11}$ $(p_{11} + q_{11})$ | $s_{10} + r_{10}$ $(p_{10} + q_{10})$ | $s_{01} + r_{01}$ $(p_{01} + q_{01})$ | $s_{00} + r_{00}$ $(p_{00} + q_{00})$ | $n$ $(1)$ |

*3.1. Hypothesis test*

Bloch (1997) studied the comparison of the weighted kappa coefficients of two *BDTs* subject to a paired design. In terms of probabilities (6) and (7), the weighted kappa coefficient of *BDT* 1 is

$$\kappa_1(c) = \frac{(p_{11} + p_{10})(q_{01} + q_{00}) - (p_{01} + p_{00})(q_{10} + q_{11})}{pc\sum_{k=0}^{1}(p_{0k} + q_{0k}) + q(1-c)\sum_{k=0}^{1}(p_{1k} + q_{1k})},$$

and that of *BDT* 2 is

$$\kappa_2(c) = \frac{(p_{11} + p_{01})(q_{10} + q_{00}) - (p_{10} + p_{00})(q_{01} + q_{11})}{pc\sum_{k=0}^{1}(p_{k0} + q_{k0}) + q(1-c)\sum_{k=0}^{1}(p_{k1} + q_{k1})}.$$

Substituting in the previous expressions the parameters by their estimators, the estimators of the weighted kappa coefficients are

$$\hat{\kappa}_1(c) = \frac{(s_{11} + s_{10})(r_{01} + r_{00}) - (s_{01} + s_{00})(r_{10} + r_{11})}{sc\sum_{k=0}^{1}(s_{0k} + r_{0k}) + r(1-c)\sum_{k=0}^{1}(s_{1k} + r_{1k})} \quad (8)$$

and

$$\hat{\kappa}_2(c) = \frac{(s_{11} + s_{01})(r_{10} + r_{00}) - (s_{10} + s_{00})(r_{01} + r_{11})}{sc\sum_{k=0}^{1}(s_{k0} + r_{k0}) + r(1-c)\sum_{k=0}^{1}(s_{k1} + r_{k1})}. \quad (9)$$

Their variances-covariance are obtained applying the delta method (see the Appendix *B* of the supplementary material). Subject to paired design the covariance between the two sensitivities and between the two specificities are given by $Cov\left[\hat{S}e_1, \hat{S}e_2\right] = \frac{\varepsilon_1}{np}$ and $Cov\left[\hat{S}p_1, \hat{S}p_2\right] = \frac{\varepsilon_0}{nq}$ respectively (Appendix B of the supplementary material), where $\varepsilon_1$ and $\varepsilon_0$ are the covariances between the two *BDTs* when $D = 1$ and $D = 0$ respectively. These covariances also affect the covariances between the two weighted kappa coefficients, just as can be seen in the expressions given in the Appendix B of the supplementary material. Finally, the statistic for the hypothesis test $H_0 : \kappa_1(c) = \kappa_2(c)$ vs $H_1 : \kappa_1(c) \neq \kappa_2(c)$ is

$$z = \frac{\hat{\kappa}_1(c) - \hat{\kappa}_2(c)}{\sqrt{\hat{V}ar\left[\hat{\kappa}_1(c)\right] + \hat{V}ar\left[\hat{\kappa}_2(c)\right] - 2\hat{C}ov\left[\hat{\kappa}_1(c), \hat{\kappa}_2(c)\right]}} \xrightarrow[n \to \infty]{} N(0,1). \quad (10)$$

*3.2. Confidence intervals*

When two parameters are compared, the interest is generally focused on studying the difference or the ratio between them. We then compare the weighted kappa coefficients of two *BDTs* through *CIs* for the difference $\delta = \kappa_1(c) - \kappa_2(c)$ and for the ratio $\theta = \kappa_1(c)/\kappa_2(c)$. Through the *CIs*: a) the two weighted kappa coefficients are compared, in such a way that if a *CI* for the difference (ratio) does not contain the zero (one) value, then we reject the equality between the weighted kappa coefficients; and b) we estimate (if the two weighted kappa coefficients are different) how much bigger one weighted kappa coefficient is than the other. Firstly, three *CIs* are proposed for the difference of the two weighted kappa coefficients, and secondly five *CIs* are proposed for the ratio.

*3.2.1. CIs for the difference*

For the difference of the two weighted kappa coefficients we propose the Wald, bootstrap and Bayesian *CIs*.

*Wald CI*. Based on the asymptotic normality of the estimator of $\delta = \kappa_1(c) - \kappa_2(c)$, i.e. $\hat{\delta} \to N[\delta, Var(\delta)]$ when the sample size $n$ is large, the Wald *CI* for the difference $\delta$ is very easy to obtain inverting the test statistic proposed by Bloch (1997), therefore

$$\delta \in \hat{\kappa}_1(c) - \hat{\kappa}_2(c) \pm z_{1-\alpha/2} \sqrt{\hat{Var}[\hat{\kappa}_1(c)] + \hat{Var}[\hat{\kappa}_2(c)] - 2\hat{Cov}[\hat{\kappa}_1(c), \hat{\kappa}_2(c)]}, \quad (11)$$

where $z_{1-\alpha/2}$ is the $100(1-\alpha/2)th$ percentile of the standard normal distribution.

*Bootstrap CI*. The bootstrap *CI* is calculated generating *B* random samples with replacement from the sample of *n* individuals. In each sample with replacement, we calculate the estimators of the weighted kappa coefficients and the difference between them, i.e. $\hat{\kappa}_{i1B}(c)$, $\hat{\kappa}_{i2B}(c)$ and $\hat{\delta}_{iB} = \hat{\kappa}_{i1B}(c) - \hat{\kappa}_{i2B}(c)$, with $i = 1, ..., B$. Then, based on the *B* differences calculated, the average difference is estimated as $\hat{\bar{\delta}}_B = \frac{1}{B}\sum_{i=1}^{B}\hat{\delta}_{iB}$.

Assuming that the bootstrap statistic $\hat{\bar{\delta}}_B$ can be transformed to a normal distribution, the bias-corrected bootstrap *CI* (Efron and Tibshirani, 1993) for $\delta$ is calculated in the following way. Let $A = \#(\hat{\delta}_{iB} < \hat{\delta})$ be the number of bootstrap estimators $\hat{\delta}_{iB}$ that are lower than the maximum likelihood estimator $\hat{\delta} = \hat{\kappa}_1(c) - \hat{\kappa}_2(c)$, and let $\hat{z}_0 = \Phi^{-1}(A/B)$, where $\Phi^{-1}(\cdot)$ is the inverse function of the standard normal cumulative distribution function. Let $\alpha_1 = \Phi(2\hat{z}_0 - z_{1-\alpha/2})$ and $\alpha_2 = \Phi(2\hat{z}_0 + z_{1-\alpha/2})$, then the bias-corrected bootstrap *CI* is $(\hat{\delta}_B^{(\alpha_1)}, \hat{\delta}_B^{(\alpha_2)})$, where $\hat{\delta}_B^{(\alpha_j)}$ is the *j*th quantile of the distribution of the *B* bootstrap estimations of $\delta$.

*Bayesian CI.* The problem is now approached from a Bayesian perspective. The number of individuals with the disease (*s*) is the product of a binomial distribution with parameters *n* and *p*, i.e. $s \to B(n,p)$. Conditioning on the individuals with the disease, i.e. conditioning on $D=1$, it is verified that

$$s_{11} + s_{10} \to B(s, Se_1) \text{ and } s_{11} + s_{01} \to B(s, Se_2). \quad (12)$$

The number of individuals without the disease (*r*) is the product of a binomial distribution with parameters *n* and *q*, i.e. $s \to B(n,q)$, with $q = 1-p$. Conditioning on the individuals without the disease $(D=0)$, it is verified that

$$r_{01} + r_{00} \to B(r, Sp_1) \text{ and } r_{10} + r_{00} \to B(r, Sp_2). \quad (13)$$

Considering the marginal distributions of each *BDT*, the estimators of the sensitivity and the specificity of the *BDT* 1, $\hat{Se}_1 = \frac{s_{11} + s_{10}}{s}$ and $\hat{Sp}_1 = \frac{r_{01} + r_{00}}{r}$, and of the *BDT* 2, $\hat{Se}_2 = \frac{s_{11} + s_{01}}{s}$ and $\hat{Sp}_2 = \frac{r_{10} + r_{00}}{r}$, are estimators of binomial proportions. In a similar way, considering the marginal distribution of the *GS*, the estimator of the disease prevalence, $\hat{p} = \frac{s}{n}$, is also the estimator of a binomial proportion. Therefore, for these estimators we propose conjugate beta prior distributions, which are the appropriate distributions for the binomial distributions involved, i.e.

$$\hat{Se}_h \to Beta(\alpha_{Se_h}, \beta_{Se_h}), \hat{Sp}_h \to Beta(\alpha_{Sp_h}, \beta_{Sp_h}) \text{ and } \hat{p} \to Beta(\alpha_p, \beta_p). \quad (14)$$

Let $\mathbf{v} = (s_{11}, s_{10}, s_{01}, s, r_{11}, r_{10}, r_{01}, n-s)$ be the vector of observed frequencies, with $s_{00} = s - s_{11} - s_{10} - s_{01}$, $r = n - s$ and $r_{00} = n - s - r_{11} - r_{10} - r_{01}$. Then the posteriori distributions for the estimators of the sensitivities, of the specificities and of the prevalence are:

$$\hat{Se}_1|\mathbf{v} \to Beta\left(s_{11} + s_{10} + \alpha_{Se_1}, s - s_{11} - s_{10} + \beta_{Se_1}\right),$$
$$\hat{Se}_2|\mathbf{v} \to Beta\left(s_{11} + s_{01} + \alpha_{Se_2}, s - s_{11} - s_{01} + \beta_{Se_2}\right),$$
$$\hat{Sp}_1|\mathbf{v} \to Beta\left(r_{01} + r_{00} + \alpha_{Sp_1}, n - s - r_{01} - r_{00} + \beta_{Sp_1}\right), \quad (15)$$
$$\hat{Sp}_2|\mathbf{v} \to Beta\left(r_{10} + r_{00} + \alpha_{Sp_2}, n - s - r_{10} - r_{00} + \beta_{Sp_2}\right),$$
$$\hat{p}|\mathbf{v} \to Beta\left(s + \alpha_p, n - s + \beta_p\right).$$

Once we have defined all distributions, the posteriori distribution for the weighted kappa coefficient of each *BDT*, and for the difference between them, can be approximated applying the Monte Carlo method. This method consists of generating *M* values of the posteriori distributions given in equations (15). In the *i*th iteration, the values generated for sensitivities $\left(\hat{Se}_h^{(i)}\right)$ and specificities $\left(\hat{Sp}_h^{(i)}\right)$ of each *BDT*, and for the prevalence $\left(\hat{p}^{(i)}\right)$, are plugged in the equations

$$\hat{\kappa}_h^{(i)}(c) = \frac{\hat{p}^{(i)}\hat{q}^{(i)}\left(\hat{Se}_h^{(i)} + \hat{Sp}_h^{(i)} - 1\right)}{\hat{p}^{(i)}\left(1 - \hat{Q}_h^{(i)}\right)c + \hat{q}^{(i)}\hat{Q}_h^{(i)}(1-c)}, \quad h = 1, 2, \quad (16)$$

where $\hat{Q}_h^{(i)} = \hat{p}^{(i)}\hat{Se}_h^{(i)} + \hat{q}^{(i)}\left(1 - \hat{Sp}_h^{(i)}\right)$. We then calculate the difference between the two weighted kappa coefficients in the *i*th iteration: $\hat{\delta}^{(i)} = \hat{\kappa}_1^{(i)}(c) - \hat{\kappa}_2^{(i)}(c)$. As the estimator of the average difference of the weighted kappa coefficients, we calculate the average of the *M* estimations of difference, i.e. $\hat{\bar{\delta}} = \frac{1}{M}\sum_{i=1}^{M}\hat{\delta}^{(i)}$. Once the Monte Carlo method is applied, based on the *M* values $\hat{\delta}^{(i)}$ we propose the calculation of a *CI* based on quantiles, i.e. the $100 \times (1-\alpha)\%$ *CI* for $\delta$ is

$$\left(q_{\alpha/2}, q_{1-\alpha/2}\right), \quad (17)$$

where $q_\gamma$ is the $\gamma$th quantile of the distribution of the *M* values $\hat{\delta}^{(i)}$.

*3.2.2. CIs for the ratio*

We propose five *CIs* for the ratio of the two weighted kappa coefficients: Wald, logarithmic, Fieller, bootstrap and Bayesian *CIs*.

*Wald CI.* Assuming the asymptotic normality of the estimator of $\theta = \kappa_1(c)/\kappa_2(c)$, i.e. $\hat{\theta} \to N[\theta, \text{Var}(\theta)]$ when the sample size *n* is large, the Wald *CI* for $\theta$ is

$$\theta \in \hat{\theta} \pm z_{1-\alpha/2}\sqrt{\hat{V}ar(\hat{\theta})}, \qquad (18)$$

where $\hat{V}ar(\hat{\theta})$ is obtained applying the delta method (Agresti, 2002), and whose expression (see Appendix B) is

$$\hat{V}ar(\hat{\theta}) \approx \frac{\hat{\kappa}_2^2(c)\hat{V}ar[\hat{\kappa}_1(c)] + \hat{\kappa}_1^2(c)\hat{V}ar[\hat{\kappa}_2(c)] - 2\hat{\kappa}_1(c)\hat{\kappa}_2(c)\hat{C}ov[\hat{\kappa}_1(c),\hat{\kappa}_2(c)]}{\hat{\kappa}_2^4(c)}.$$

Expressions of the variances-covariance can be seen in Appendix B.

*Logarithmic CI.* Assuming the asymptotic normality of the Napierian logarithm of the $\hat{\theta}$, i.e. $\ln(\hat{\theta}) \to N(\ln(\theta), \text{Var}[\ln(\theta)])$ when the sample size *n* is large, an asymptotic *CI* for $\ln(\theta)$ is

$$\ln(\theta) \in \ln(\hat{\theta}) \pm z_{1-\alpha/2}\sqrt{\hat{V}ar[\ln(\hat{\theta})]}.$$

Taking exponential, the logarithmic *CI* for $\theta$ is

$$\theta \in \hat{\theta} \times \exp\left\{\pm z_{1-\alpha/2}\sqrt{\hat{V}ar[\ln(\hat{\theta})]}\right\}, \qquad (19)$$

where $\hat{V}ar[\ln(\hat{\theta})]$ is obtained applying the delta method (see Appendix B), i.e.

$$\hat{V}ar[\ln(\hat{\theta})] \approx \frac{\hat{V}ar[\hat{\kappa}_1(c)]}{\hat{\kappa}_1^2(c)} + \frac{\hat{V}ar[\hat{\kappa}_2(c)]}{\hat{\kappa}_2^2(c)} - \frac{2\hat{C}ov[\hat{\kappa}_1(c),\hat{\kappa}_2(c)]}{\hat{\kappa}_1(c)\hat{\kappa}_2(c)}.$$

*Fieller CI.* The Fieller method (1940) is a classic method to obtain a *CI* for the ratio of two parameters. This method requires us to assume that the estimators are distributed according to a normal bivariate distribution, i.e. $\left(\hat{\kappa}_1(c), \hat{\kappa}_2(c)\right)^T \to N\left[\boldsymbol{\kappa}(c), \Sigma_{\boldsymbol{\kappa}(c)}\right]$ when the sample size $n$ is large, where $\boldsymbol{\kappa}(c) = \left(\kappa_1(c), \kappa_2(c)\right)^T$ and

$$\Sigma_{\boldsymbol{\kappa}(c)} = \begin{pmatrix} \sigma_{11} & \sigma_{12} \\ \sigma_{21} & \sigma_{22} \end{pmatrix} = \begin{pmatrix} Var\left[\kappa_1(c)\right] & Cov\left[\kappa_1(c), \kappa_2(c)\right] \\ Cov\left[\kappa_1(c), \kappa_2(c)\right] & Var\left[\kappa_2(c)\right] \end{pmatrix}.$$

Applying the Fieller method it is verified that $\hat{\kappa}_1(c) - \theta\hat{\kappa}_2(c) \xrightarrow[n\to\infty]{} N\left(0, \sigma_{11} - 2\theta\sigma_{12} + \theta^2\sigma_{22}\right)$. The Fieller CI is obtained by searching for the set of values for $\theta$ that satisfy the inequality

$$\frac{\left[\hat{\kappa}_1(c) - \theta\hat{\kappa}_2(c)\right]^2}{\hat{\sigma}_{11} - 2\theta\hat{\sigma}_{12} + \theta^2\hat{\sigma}_{22}} < z_{1-\alpha/2}^2.$$

Finally, the Fieller CI for $\theta = \kappa_1(c)/\kappa_2(c)$ is

$$\theta \in \frac{\hat{\omega}_{12} \pm \sqrt{\hat{\omega}_{12}^2 - \hat{\omega}_{11}\hat{\omega}_{22}}}{\hat{\omega}_{22}}, \tag{20}$$

where $\hat{\omega}_{ij} = \hat{\kappa}_i(c) \times \hat{\kappa}_j(c) - \hat{\sigma}_{ij} z_{1-\alpha/2}^2$ with $i, j = 1, 2$, and verifying that $\hat{\omega}_{12} = \hat{\omega}_{21}$. This interval is valid when $\hat{\omega}_{12}^2 > \hat{\omega}_{11}\hat{\omega}_{22}$ and $\hat{\omega}_{22} \neq 0$.

*Bootstrap CI*. The bootstrap CI for $\theta$ is calculated in a similar way to that of the bootstrap interval explained in Section 3.1 but considering $\theta$ instead of $\delta$. In each sample with replacement obtained we calculate the estimators of the weighted kappa coefficients and the ratio between them, i.e. $\hat{\kappa}_{i1B}(c)$, $\hat{\kappa}_{i2B}(c)$ and $\hat{\theta}_{iB} = \hat{\kappa}_{i1B}(c)/\hat{\kappa}_{i2B}(c)$, with $i = 1, ..., B$. Then, based on the $B$ ratios calculated we estimate the average ratio as $\hat{\bar{\theta}}_B = \frac{1}{B}\sum_{i=1}^{B}\hat{\theta}_{iB}$. Assuming that the statistic $\hat{\bar{\theta}}_B$ can be transformed to a normal distribution, the bias-corrected bootstrap CI (Efron and Tibshirani, 1993) for $\theta$ is obtained in a similar way to how the bootstrap CI for $\delta$ is calculated, considering now that $A = \#\left(\hat{\theta}_{iB} < \hat{\theta}\right)$. Finally, the bias-corrected bootstrap *CI* is $\left(\hat{\theta}_B^{(\alpha_1)}, \hat{\theta}_B^{(\alpha_2)}\right)$, where $\hat{\theta}_B^{(\alpha_j)}$ is the *j*th quantile of the distribution of the *B* bootstrap estimations of $\theta$.

*Bayesian CI*. The Bayesian *CI* for $\theta$ is also calculated in a similar way to that of the bayesian *CI* presented in Section 3.1. Considering the same distributions given in equations (14) and (15), in the *i*th iteration of the Monte Carlo method we calculate the ratio $\hat{\theta}^{(i)} = \hat{\kappa}_1^{(i)}(c) / \hat{\kappa}_2^{(i)}(c)$ and as an estimator we calculate $\hat{\bar{\theta}} = \frac{1}{M} \sum_{i=1}^{M} \hat{\theta}^{(i)}$. Finally, based on the *M* values $\hat{\theta}^{(i)}$ we calculate the *CI* based on quantiles.

The five previous *CIs* are for the ratio $\theta = \kappa_1(c) / \kappa_2(c)$. If we want to calculate the *CI* for the ratio $\kappa_2(c) / \kappa_1(c)$ $(= \theta' = 1/\theta)$, then the logarithmic, Fieller, bootstrap and Bayesian *CIs* are obtained by calculating the inverse of each boundary of the corresponding *CI* for $\theta = \kappa_1(c) / \kappa_2(c)$. Nevertheless, the Wald *CI* for $\theta'$ is obtained from the Wald *CI* for $\theta$ dividing each boundary by $\hat{\theta}^2$, i.e. if $(L_\theta, U_\theta)$ is the Wald *CI* for $\theta = \kappa_1(c) / \kappa_2(c)$ then the Wald *CI* for $\theta' = \kappa_2(c) / \kappa_1(c)$ is $(L_\theta / \hat{\theta}^2, U_\theta / \hat{\theta}^2)$.

## 4. Simulation experiments

Monte Carlo simulation experiments were carried out to study the coverage probability (*CP*) and the average length (*AL*) of each of the *CIs* presented in Section 3.2. For this purpose, we generated $N = 10,000$ random samples with multinomial distribution sized $n = \{25, 50, 100, 200, 300, 400, 500, 1000\}$. The random samples were generated setting the values of the weighted kappa coefficients, following these steps:

1. For the disease prevalence, we took the values $p = \{5\%, 10\%, 25\%, 50\%\}$.

2. For the weighting index, we took a small, intermediate and high value: $c = \{0.1, 0.5, 0.9\}$.

3. As values of the weighted kappa coefficients with $c = 0$ and $c = 1$, we took the following values: $\kappa_h(0), \kappa_h(1) = \{0.01, 0.02, ..., 0.98, 0.99\}$.

4. Next, using all of the values set previously, we calculated the sensitivity and the specificity of each diagnostic test solving the equations

$$Se_h = \frac{[q\kappa_h(0) + p]\kappa_h(1)}{q\kappa_h(0) + p\kappa_h(1)} \text{ and } Se_h = \frac{[p\kappa_h(1) + q]\kappa_h(0)}{q\kappa_h(0) + p\kappa_h(1)},$$

considering, quite logically, only those cases in which the Youden index is higher than 0, i.e. $Y_h = Se_h + Sp_h - 1 > 0$.

5. The values of $\kappa_h(c)$ were calculated applying the equation

$$\kappa_h(c) = \frac{pc(1-Q_h)\kappa_h(1) + q(1-c)Q_h\kappa_h(0)}{pc(1-Q_h) + q(1-c)Q_h},$$

where $Q_h = pSe_h + q(1-Sp_h)$.

6. As values of the weighted kappa coefficients we considered $\kappa_h(c) = \{0.2, 0.4, 0.6, 0.8\}$, and from these we calculated $\delta$ and $\theta$. In order to be able to compare the coverage probabilities of the *CIs* for $\delta$ and for $\theta$, $\kappa_1(c)$ and $\kappa_2(c)$ must be the same for $\delta$ and $\theta$.

Following the idea of Cicchetti (2001), simulations were carried out for values of $\kappa_h(c)$ with different levels of significance: poor $(\kappa_h(c) < 0.40)$, fair $(0.40 \leq \kappa_h(c) \leq 0.59)$, good $(0.60 \leq \kappa_h(c) \leq 0.74)$ and excellent $(0.75 \leq \kappa_h(c) \leq 1)$. As values of the dependence factors $\varepsilon_1$ and $\varepsilon_0$ we took intermediate values (50% of the maximum value of each $\varepsilon_i$) and high values (80% of the maximum value of each $\varepsilon_i$), i.e.

$$\varepsilon_1 = f \times Min\{Se_1(1-Se_2), Se_2(1-Se_1)\} \quad \text{and} \quad \varepsilon_0 = f \times Min\{Sp_1(1-Sp_2), Sp_2(1-Sp_1)\}$$

where $f = \{0.50, 0.80\}$. Probabilities of the multinomial distributions, equations (6) and (7), were calculated from values of the weighted kappa coefficients, and not setting the values of the sensitivities and specificities. In each scenario considered, for each one of the $N$ random samples we calculated all the *CIs* proposed in Section 3.2. For the bayesian *CIs* we considered as prior distribution a $Beta(1,1)$ distribution for all of the estimators (sensitivities, specificities and prevalence). This distribution is a non-informative distribution and is flat for all possible values of each sensitivity, specificity and prevalence, and has a minimum impact on each posteriori distribution. For the bootstrap method, for each one of the $N$ random samples we also generated $B = 2,000$ samples with replacement; and for the Bayesian method, for each one of the $N$ random samples we also generated another $M = 10,000$. Moreover, the simulation experiments were designed in such a way that in all of the random samples generated we can estimate the weighted kappa coefficients and their variances-covariance, in order to be able to calculate all of the intervals proposed in Section 3.2. As the confidence level, we took 95%.

The comparison of the asymptotic behaviour of the *CIs* was made following a similar procedure to that used by other authors (Price and Bonett, 2004; Martín-Andrés and Álvarez-Hernández, 2014a, 2014b; Montero-Alonso and Roldán-Nofuentes, 2019). This procedure consists of determining if the *CI* "fails" for a confidence of 95%, which happens if the *CI* has a $CP \leq 93\%$. The selection of the *CI* with the best asymptotic behaviour (for the difference and for the ratio) was made following the following steps: 1) Choose the *CIs* with the least failures ($CP > 93\%$), and 2) Choose the *CIs* which are the most accurate, i.e. those which have the lowest *AL*. In the Appendix *C* of the supplementary material this method is justified.

### 4.1. CIs for the difference $\delta$

Tables 3 and 4 show some of the results obtained (*CPs* and *ALs*) for $\delta = \{-0.6, -0.4, -0.2, 0\}$, indicating in each case the scenarios ($\kappa_h(c)$, $Se_h$, $Sp_h$ and $p$) in which these values were obtained, and for intermediate values of the dependence factors $\varepsilon_1$ and $\varepsilon_0$. These Tables indicate the failures in bold type and it was considered that $\kappa_1(c) \leq \kappa_2(c)$. If it is considered that $\kappa_1(c) > \kappa_2(c)$, the *CPs* are the same and the conclusions too. From the results, the following conclusions are obtained:

a) Wald *CI*. For $\delta = \{-0.6, -0.4\}$ the Wald *CI* fails for a small $(n \leq 50)$ and a moderate $(n = 100)$ sample size, and for a large sample size $(n \geq 200)$ the Wald *CI* does not fail. For $\delta = \{-0.2, 0\}$ the Wald *CI* does not fail.

b) Bootstrap *CI*. In very general terms, for $\delta = \{-0.6, -0.4\}$ this *CI* fails when $n \leq 100$, and for $n \geq 200$ this interval does not fail. For $\delta = -0.2$ this *CI* fails for almost all the sample sizes, and for $\delta = 0$ does not fail. When this *CI* does not fail, the *AL* is slightly lower than the Wald *CI* for $\delta = \{-0.2, 0\}$, and slightly higher for $\delta = \{-0.6, -0.4\}$ and $n \geq 200$.

c) Bayesian *CI*. In very general terms, for $\delta = \{-0.6, -0.4\}$ this *CI* fails when $n \leq 50$, whereas for $n \geq 100$ this *CI* does not fail. For $\delta = \{-0.2, 0\}$ this *CI* does not fail. Regarding the *AL*, in the situations in which it does not fail, the *AL* is slightly higher than the *ALs* of the Wald *CI* and of the bootstrap *CI*.

Table 3. Coverage probabilities (*CPs*) and average lengths (*ALs*) of the *CIs* for the difference $\delta$ of the two weighted kappa coefficients (I).

$\kappa_1(0.1)=0.2\ \kappa_2(0.1)=0.8\ \delta=-0.6$
$Se_1=0.484\ Sp_1=0.684\ Se_2=0.852\ Sp_2=0.911\ \varepsilon_1=0.0359\ \varepsilon_0=0.0306\ p=50\%$

| | Wald *CI* | | Bootstrap *CI* | | Bayesian *CI* | |
|---|---|---|---|---|---|---|
| *n* | CP | AL | CP | AL | CP | AL |
| 25 | **0.335** | 0.866 | **0** | 0.643 | **0.287** | 0.923 |
| 50 | **0.737** | 0.646 | **0.038** | 0.589 | **0.762** | 0.690 |
| 100 | **0.912** | 0.470 | **0.750** | 0.473 | 0.937 | 0.501 |
| 200 | 0.958 | 0.337 | 0.952 | 0.354 | 0.968 | 0.364 |
| 300 | 0.972 | 0.276 | 0.980 | 0.295 | 0.982 | 0.301 |
| 400 | 0.960 | 0.239 | 0.969 | 0.258 | 0.971 | 0.262 |
| 500 | 0.955 | 0.214 | 0.972 | 0.231 | 0.975 | 0.236 |
| 1000 | 0.937 | 0.152 | 0.963 | 0.164 | 0.965 | 0.168 |

$\kappa_1(0.9)=0.2\ \kappa_2(0.9)=0.8\ \delta=-0.6$
$Se_1=0.28\ Sp_1=0.92\ Se_2=0.82\ Sp_2=0.98\ \varepsilon_1=0.0252\ \varepsilon_0=0.0092\ p=10\%$

| | Wald *CI* | | Bootstrap *CI* | | Bayesian *CI* | |
|---|---|---|---|---|---|---|
| *n* | CP | AL | CP | AL | CP | AL |
| 25 | **0.114** | 0.999 | **0** | 0.651 | **0.033** | 0.987 |
| 50 | **0.566** | 0.863 | **0** | 0.640 | **0.280** | 0.838 |
| 100 | **0.760** | 0.682 | **0.031** | 0.614 | **0.600** | 0.667 |
| 200 | **0.885** | 0.503 | **0.487** | 0.490 | **0.815** | 0.503 |
| 300 | 0.934 | 0.411 | **0.733** | 0.402 | **0.886** | 0.418 |
| 400 | 0.935 | 0.354 | **0.823** | 0.347 | **0.903** | 0.365 |
| 500 | 0.947 | 0.314 | **0.892** | 0.309 | 0.937 | 0.326 |
| 1000 | 0.947 | 0.220 | 0.938 | 0.218 | 0.947 | 0.233 |

$\kappa_1(0.1)=0.4\ \kappa_2(0.1)=0.8\ \delta=-0.4$
$Se_1=0.804\ Sp_1=0.887\ Se_2=0.82\ Sp_2=0.98\ \varepsilon_1=0.0723\ \varepsilon_0=0.0089\ p=10\%$

| | Wald *CI* | | Bootstrap *CI* | | Bayesian *CI* | |
|---|---|---|---|---|---|---|
| *n* | CP | AL | CP | AL | CP | AL |
| 25 | **0.847** | 0.812 | **0.473** | 0.671 | **0.920** | 0.899 |
| 50 | **0.856** | 0.715 | **0.602** | 0.608 | **0.910** | 0.764 |
| 100 | **0.924** | 0.534 | **0.847** | 0.528 | 0.953 | 0.580 |
| 200 | 0.968 | 0.373 | 0.955 | 0.423 | 0.978 | 0.426 |
| 300 | 0.957 | 0.302 | 0.986 | 0.367 | 0.976 | 0.369 |
| 400 | 0.951 | 0.261 | 0.992 | 0.313 | 0.978 | 0.315 |
| 500 | 0.955 | 0.232 | 0.994 | 0.259 | 0.979 | 0.262 |
| 1000 | 0.941 | 0.164 | 0.994 | 0.202 | 0.967 | 0.204 |

$\kappa_1(0.5)=0.4\ \kappa_2(0.5)=0.8\ \delta=-0.4$
$Se_1=0.76\ Sp_1=0.72\ Se_2=0.85\ Sp_2=0.95\ \varepsilon_1=0.0570\ \varepsilon_0=0.0180\ p=25\%$

| | Wald *CI* | | Bootstrap *CI* | | Bayesian *CI* | |
|---|---|---|---|---|---|---|
| *n* | CP | AL | CP | AL | CP | AL |
| 25 | **0.894** | 0.810 | **0.004** | 0.613 | 0.962 | 0.858 |
| 50 | 0.935 | 0.580 | **0.516** | 0.516 | 0.961 | 0.641 |
| 100 | 0.945 | 0.397 | **0.824** | 0.379 | 0.970 | 0.458 |
| 200 | 0.946 | 0.275 | **0.928** | 0.271 | 0.971 | 0.320 |
| 300 | 0.952 | 0.221 | 0.934 | 0.220 | 0.974 | 0.259 |
| 400 | 0.940 | 0.191 | 0.938 | 0.192 | 0.963 | 0.224 |
| 500 | 0.948 | 0.171 | 0.942 | 0.170 | 0.979 | 0.200 |
| 1000 | 0.945 | 0.120 | 0.944 | 0.119 | 0.979 | 0.140 |

Similar conclusions are obtained when the dependence factors take high values. Therefore, regarding the effect of the dependence factors $\varepsilon_i$ on the asymptotic behaviour of the *CIs*, in general terms they do not have a clear effect on the *CPs* of the *CIs*.

Table 4. Coverage probabilities (*CPs*) and average lengths (*ALs*) of the *CIs* for the difference $\delta$ of the two weighted kappa coefficients (II).

$\kappa_1(0.9)=0.6 \quad \kappa_2(0.9)=0.8 \quad \delta=-0.2$

$Se_1=0.62 \quad Sp_1=0.98 \quad Se_2=0.911 \quad Sp_2=0.937 \quad \varepsilon_1=0.0277 \quad \varepsilon_0=0.0094 \quad p=5\%$

|   | Wald *CI* | | Bootstrap *CI* | | Bayesian *CI* | |
|---|---|---|---|---|---|---|
| n | CP | AL | CP | AL | CP | AL |
| 25 | 1 | 1.009 | **0.757** | 0.724 | 1 | 1.018 |
| 50 | 0.996 | 0.913 | **0.829** | 0.659 | 0.999 | 0.916 |
| 100 | 0.993 | 0.823 | **0.928** | 0.580 | 0.998 | 0.801 |
| 200 | 0.934 | 0.642 | **0.763** | 0.535 | 0.986 | 0.649 |
| 300 | 0.922 | 0.533 | **0.745** | 0.483 | 0.964 | 0.551 |
| 400 | 0.941 | 0.456 | **0.794** | 0.434 | 0.971 | 0.481 |
| 500 | 0.933 | 0.404 | **0.799** | 0.393 | 0.962 | 0.430 |
| 1000 | 0.948 | 0.282 | **0.913** | 0.282 | 0.967 | 0.305 |

$\kappa_1(0.1)=0.6 \quad \kappa_2(0.1)=0.8 \quad \delta=-0.2$

$Se_1=0.195 \quad Sp_1=0.995 \quad Se_2=0.477 \quad Sp_2=0.987 \quad \varepsilon_1=0.0509 \quad \varepsilon_0=0.0026 \quad p=25\%$

|   | Wald *CI* | | Bootstrap *CI* | | Bayesian *CI* | |
|---|---|---|---|---|---|---|
| n | CP | AL | CP | AL | CP | AL |
| 25 | 1 | 0.928 | 1 | 0.644 | 1 | 0.981 |
| 50 | 0.999 | 0.787 | 1 | 0.613 | 1 | 0.866 |
| 100 | 0.994 | 0.604 | 0.999 | 0.581 | 0.999 | 0.692 |
| 200 | 0.985 | 0.429 | 0.997 | 0.464 | 0.998 | 0.505 |
| 300 | 0.981 | 0.347 | 0.991 | 0.393 | 0.994 | 0.411 |
| 400 | 0.973 | 0.297 | 0.986 | 0.346 | 0.992 | 0.352 |
| 500 | 0.967 | 0.263 | 0.984 | 0.311 | 0.989 | 0.311 |
| 1000 | 0.957 | 0.182 | 0.988 | 0.222 | 0.987 | 0.213 |

$\kappa_1(0.5)=0.4 \quad \kappa_2(0.5)=0.4 \quad \delta=0$

$Se_1=0.76 \quad Sp_1=0.72 \quad Se_2=0.40 \quad Sp_2=0.943 \quad \varepsilon_1=0.0480 \quad \varepsilon_0=0.0206 \quad p=25\%$

|   | Wald *CI* | | Bootstrap *CI* | | Bayesian *CI* | |
|---|---|---|---|---|---|---|
| n | CP | AL | CP | AL | CP | AL |
| 25 | 0.990 | 0.811 | 0.988 | 0.624 | 0.999 | 0.826 |
| 50 | 0.978 | 0.683 | 0.998 | 0.598 | 0.994 | 0.691 |
| 100 | 0.962 | 0.499 | 0.967 | 0.466 | 0.985 | 0.522 |
| 200 | 0.955 | 0.353 | 0.963 | 0.340 | 0.981 | 0.381 |
| 300 | 0.944 | 0.288 | 0.943 | 0.280 | 0.965 | 0.314 |
| 400 | 0.960 | 0.250 | 0.962 | 0.244 | 0.980 | 0.274 |
| 500 | 0.946 | 0.223 | 0.945 | 0.219 | 0.966 | 0.246 |
| 1000 | 0.951 | 0.158 | 0.951 | 0.155 | 0.972 | 0.175 |

$\kappa_1(0.9)=0.4 \quad \kappa_2(0.9)=0.4 \quad \delta=0$

$Se_1=0.943 \quad Sp_1=0.229 \quad Se_2=0.70 \quad Sp_2=0.70 \quad \varepsilon_1=0.0200 \quad \varepsilon_0=0.0343 \quad p=50\%$

|   | Wald *CI* | | Bootstrap *CI* | | Bayesian *CI* | |
|---|---|---|---|---|---|---|
| n | CP | AL | CP | AL | CP | AL |
| 25 | 1 | 0.936 | 1 | 0.735 | 1 | 0.950 |
| 50 | 0.997 | 0.788 | 0.997 | 0.717 | 1 | 0.786 |
| 100 | 0.992 | 0.602 | 0.982 | 0.578 | 0.997 | 0.617 |
| 200 | 0.980 | 0.435 | 0.981 | 0.432 | 0.990 | 0.461 |
| 300 | 0.959 | 0.356 | 0.965 | 0.358 | 0.973 | 0.382 |
| 400 | 0.951 | 0.307 | 0.958 | 0.311 | 0.972 | 0.332 |
| 500 | 0.956 | 0.274 | 0.958 | 0.278 | 0.969 | 0.297 |
| 1000 | 0.956 | 0.193 | 0.958 | 0.196 | 0.970 | 0.210 |

## 4.2. *CIs* for the ratio $\theta$

Tables 5 and 6 show some of the results obtained for $\theta=\{0.25, 0.50, 0.75, 1\}$, considering the same scenarios as in Tables 3 and 4. As in the case of the previous *CIs*, it was considered that $\kappa_1(c) \leq \kappa_2(c)$, and the same conclusions are obtained if $\kappa_1(c) > \kappa_2(c)$. From the results, the following conclusions are obtained:

a) Wald *CI*. The Wald *CI* fails when $\theta = 0.25$ and the sample size is small $(n \leq 50)$ or moderate $(n = 100)$, and this *CI* does not fail for the rest of the values of $\theta$ and sample sizes.

b) Logarithmic *CI*. This *CI* fails when $\theta = \{0.25, 0.50\}$ and $n \leq 200 - 300$ depending on the value of $\theta$. For $\theta = 0.75$ this *CI* fails for some large sample sizes, and for $\theta = 1$ it does not fail. This *CI* fails more than the Wald *CI*, and in the situations in which it does not fail, its *AL* is slightly higher than that of the Wald *CI*.

c) Fieller *CI*. This *CI* fails when $\theta = \{0.25, 0.5\}$ and $n \leq 50$, and it does not fail for the rest of the values of $\theta$ and sample sizes. In general terms, when there are no failures, its *AL* is similar to that of the Wald and Logarithmic *CI*s.

d) Bootstrap *CI*. This *CI* has numerous failures when $\theta = \{0.25, 0.50, 0.75\}$, whereas for $\theta = 1$ it does not fail. When $\theta = 1$, its *AL* is greater than that of the Wald and Logarithmic *CI*s, especially when $n \leq 400$, and its *AL* is also slightly lower than that of the Fieller *CI*.

e) Bayesian *CI*. This *CI* only fails when $\theta = 0.25$ and $n \leq 50$. When this *CI* does not fail, its *AL* is, in general terms, somewhat larger than that of the rest of the *CI*s.

Table 5. Coverage probabilities (*CPs*) and average lengths (*ALs*) of the *CIs* for the ratio $\theta$ of the two weighted kappa coefficients (I).

$\kappa_1(0.1) = 0.2$  $\kappa_2(0.1) = 0.8$  $\theta = 0.25$

$Se_1 = 0.484$  $Sp_1 = 0.684$  $Se_2 = 0.852$  $Sp_2 = 0.911$  $\varepsilon_1 = 0.0359$  $\varepsilon_0 = 0.0306$  $p = 50\%$

| | Wald *CI* | | Logarit. *CI* | | Fieller *CI* | | Bootstrap *CI* | | Bayesian *CI* | |
|---|---|---|---|---|---|---|---|---|---|---|
| *n* | CP | AL | CP | AL | CP | AL | CP | AL | CP | AL |
| 25 | **0.823** | 1.351 | **0.088** | 1.517 | **0.700** | 1.950 | **0.368** | 2.260 | **0.884** | 2.704 |
| 50 | **0.837** | 0.803 | **0.532** | 0.886 | **0.828** | 0.851 | **0.634** | 0.882 | **0.905** | 0.965 |
| 100 | 0.931 | 0.551 | **0.832** | 0.608 | 0.942 | 0.565 | **0.889** | 0.569 | 0.954 | 0.585 |
| 200 | 0.957 | 0.389 | **0.920** | 0.422 | 0.962 | 0.392 | 0.952 | 0.388 | 0.970 | 0.402 |
| 300 | 0.970 | 0.318 | 0.933 | 0.340 | 0.974 | 0.319 | 0.969 | 0.316 | 0.984 | 0.328 |
| 400 | 0.960 | 0.277 | 0.936 | 0.293 | 0.967 | 0.278 | 0.962 | 0.276 | 0.976 | 0.285 |
| 500 | 0.957 | 0.248 | 0.944 | 0.260 | 0.967 | 0.248 | 0.969 | 0.247 | 0.975 | 0.256 |
| 1000 | 0.945 | 0.175 | 0.963 | 0.179 | 0.944 | 0.176 | 0.943 | 0.175 | 0.953 | 0.182 |

$\kappa_1(0.9) = 0.2$  $\kappa_2(0.9) = 0.8$  $\theta = 0.25$

$Se_1 = 0.28$  $Sp_1 = 0.92$  $Se_2 = 0.82$  $Sp_2 = 0.98$  $\varepsilon_1 = 0.0252$  $\varepsilon_0 = 0.00092$  $p = 10\%$

| | Wald *CI* | | Logarit. *CI* | | Fieller *CI* | | Bootstrap *CI* | | Bayesian *CI* | |
|---|---|---|---|---|---|---|---|---|---|---|
| *n* | CP | AL | CP | AL | CP | AL | CP | AL | CP | AL |
| 25 | **0.885** | 1.760 | **0.002** | 2.029 | **0.566** | 3.567 | **0.011** | 3.175 | **0.866** | 3.851 |
| 50 | **0.916** | 1.249 | **0.259** | 1.415 | **0.765** | 1.660 | **0.040** | 1.722 | **0.767** | 1.816 |
| 100 | 0.936 | 0.846 | **0.636** | 0.947 | **0.884** | 0.939 | **0.363** | 1.048 | **0.843** | 0.986 |
| 200 | 0.958 | 0.560 | **0.835** | 0.617 | 0.945 | 0.581 | **0.807** | 0.607 | 0.932 | 0.594 |
| 300 | 0.967 | 0.440 | **0.900** | 0.479 | 0.960 | 0.450 | **0.902** | 0.456 | 0.948 | 0.459 |
| 400 | 0.965 | 0.373 | 0.931 | 0.402 | 0.959 | 0.379 | 0.932 | 0.380 | 0.943 | 0.387 |
| 500 | 0.971 | 0.327 | 0.936 | 0.349 | 0.971 | 0.331 | 0.942 | 0.330 | 0.960 | 0.339 |
| 1000 | 0.950 | 0.227 | 0.941 | 0.235 | 0.950 | 0.228 | 0.949 | 0.227 | 0.955 | 0.234 |

$\kappa_1(0.1) = 0.4$  $\kappa_2(0.1) = 0.8$  $\theta = 0.5$

$Se_1 = 0.804$  $Sp_1 = 0.887$  $Se_2 = 0.82$  $Sp_2 = 0.98$  $p = 10\%$

$\varepsilon_1 = 0.0723$  $\varepsilon_0 = 0.0089$

| | Wald *CI* | | Logarit. *CI* | | Fieller *CI* | | Bootstrap *CI* | | Bayesian *CI* | |
|---|---|---|---|---|---|---|---|---|---|---|
| *n* | CP | AL | CP | AL | CP | AL | CP | AL | CP | AL |
| 25 | **0.918** | 1.141 | **0.835** | 1.259 | **0.893** | 2.824 | **0.543** | 1.157 | **0.906** | 2.310 |
| 50 | 0.959 | 1.021 | **0.859** | 1.119 | 0.939 | 1.518 | **0.897** | 1.140 | 0.978 | 1.710 |
| 100 | 0.961 | 0.619 | **0.922** | 0.655 | 0.949 | 0.693 | **0.880** | 0.670 | 0.975 | 0.828 |
| 200 | 0.962 | 0.395 | 0.947 | 0.406 | 0.959 | 0.409 | **0.914** | 0.400 | 0.977 | 0.470 |
| 300 | 0.955 | 0.315 | 0.951 | 0.320 | 0.956 | 0.321 | **0.928** | 0.312 | 0.976 | 0.363 |
| 400 | 0.953 | 0.271 | 0.949 | 0.274 | 0.952 | 0.274 | 0.935 | 0.265 | 0.975 | 0.308 |
| 500 | 0.951 | 0.240 | 0.950 | 0.242 | 0.953 | 0.242 | 0.932 | 0.234 | 0.971 | 0.271 |
| 1000 | 0.939 | 0.169 | 0.943 | 0.170 | 0.939 | 0.170 | 0.934 | 0.163 | 0.963 | 0.189 |

$\kappa_1(0.5) = 0.4$  $\kappa_2(0.5) = 0.8$  $\theta = 0.5$

$Se_1 = 0.76$  $Sp_1 = 0.72$  $Se_2 = 0.85$  $Sp_2 = 0.95$  $\varepsilon_1 = 0.0570$  $\varepsilon_0 = 0.0180$  $p = 25\%$

| | Wald *CI* | | Logarit. *CI* | | Fieller *CI* | | Bootstrap *CI* | | Bayesian *CI* | |
|---|---|---|---|---|---|---|---|---|---|---|
| *n* | CP | AL | CP | AL | CP | AL | CP | AL | CP | AL |
| 25 | 0.997 | 1.328 | **0.918** | 1.493 | 0.966 | 2.222 | **0.901** | 2.463 | 0.999 | 2.825 |
| 50 | 0.983 | 0.780 | **0.924** | 0.848 | 0.966 | 0.855 | **0.925** | 0.894 | 0.995 | 1.057 |
| 100 | 0.977 | 0.488 | 0.957 | 0.510 | 0.969 | 0.501 | 0.952 | 0.498 | 0.990 | 0.586 |
| 200 | 0.958 | 0.323 | 0.956 | 0.329 | 0.957 | 0.327 | 0.940 | 0.320 | 0.981 | 0.372 |
| 300 | 0.958 | 0.257 | 0.954 | 0.260 | 0.957 | 0.259 | 0.945 | 0.252 | 0.978 | 0.292 |
| 400 | 0.948 | 0.221 | 0.947 | 0.222 | 0.948 | 0.221 | 0.936 | 0.215 | 0.966 | 0.249 |
| 500 | 0.954 | 0.196 | 0.953 | 0.197 | 0.954 | 0.196 | 0.943 | 0.190 | 0.972 | 0.220 |
| 1000 | 0.944 | 0.137 | 0.951 | 0.137 | 0.945 | 0.137 | 0.933 | 0.132 | 0.968 | 0.152 |

Table 6. Coverage probabilities (*CPs*) and average lengths (*ALs*) of the *CIs* for the ratio $\theta$ of the two weighted kappa coefficients (II).

$\kappa_1(0.9)=0.6 \quad \kappa_2(0.9)=0.8 \quad \theta=0.75$

$Se_1=0.62 \quad Sp_1=0.98 \quad Se_2=0.911 \quad Sp_2=0.936 \quad \varepsilon_1=0.0277 \quad \varepsilon_0=0.0094 \quad p=5\%$

|   | Wald *CI* | | Logarit. *CI* | | Fieller *CI* | | Bootstrap *CI* | | Bayesian *CI* | |
|---|---|---|---|---|---|---|---|---|---|---|
| *n* | *CP* | *AL* | *CP* | *AL* | *CP* | *AL* | *CP* | *AL* | *CP* | *AL* |
| 25 | 1 | 1.514 | 1 | 1.679 | 1 | 2.689 | 0.999 | 2.578 | 1 | 3.538 |
| 50 | 0.999 | 1.409 | 0.994 | 1.487 | 0.993 | 1.972 | 0.979 | 2.311 | 1 | 2.392 |
| 100 | 0.999 | 1.323 | 0.993 | 1.451 | 0.993 | 1.899 | 0.975 | 1.425 | 1 | 1.980 |
| 200 | 0.971 | 0.909 | 0.933 | 0.965 | 0.940 | 1.037 | 0.965 | 0.998 | 0.991 | 1.173 |
| 300 | 0.946 | 0.709 | 0.916 | 0.738 | 0.939 | 0.767 | 0.958 | 0.784 | 0.973 | 0.854 |
| 400 | 0.955 | 0.583 | 0.933 | 0.599 | 0.944 | 0.601 | 0.959 | 0.620 | 0.977 | 0.679 |
| 500 | 0.943 | 0.506 | 0.925 | 0.516 | 0.931 | 0.516 | 0.961 | 0.551 | 0.969 | 0.579 |
| 1000 | 0.947 | 0.341 | 0.945 | 0.344 | 0.943 | 0.344 | 0.969 | 0.375 | 0.969 | 0.377 |

$\kappa_1(0.1)=0.6 \quad \kappa_2(0.1)=0.8 \quad \theta=0.75$

$Se_1=0.195 \quad Sp_1=0.995 \quad Se_2=0.477 \quad Sp_2=0.987 \quad \varepsilon_1=0.0509 \quad \varepsilon_0=0.0026 \quad p=25\%$

|   | Wald *CI* | | Logarit. *CI* | | Fieller *CI* | | Bootstrap *CI* | | Bayesian *CI* | |
|---|---|---|---|---|---|---|---|---|---|---|
| *n* | *CP* | *AL* | *CP* | *AL* | *CP* | *AL* | *CP* | *AL* | *CP* | *AL* |
| 25 | 1 | 1.687 | 1 | 1.924 | 1 | 4.747 | 1 | 2.676 | 1 | 4.561 |
| 50 | 1 | 1.266 | 1 | 1.400 | 1 | 2.837 | 1 | 1.609 | 1 | 2.308 |
| 100 | 0.999 | 0.865 | 0.997 | 0.923 | 0.997 | 0.946 | 0.998 | 0.945 | 1 | 1.188 |
| 200 | 0.992 | 0.565 | 0.990 | 0.583 | 0.986 | 0.579 | 0.975 | 0.618 | 0.997 | 0.700 |
| 300 | 0.971 | 0.444 | 0.990 | 0.452 | 0.976 | 0.449 | 0.958 | 0.493 | 0.992 | 0.536 |
| 400 | 0.971 | 0.375 | 0.985 | 0.380 | 0.972 | 0.378 | 0.960 | 0.420 | 0.989 | 0.448 |
| 500 | 0.966 | 0.328 | 0.976 | 0.331 | 0.971 | 0.331 | 0.964 | 0.371 | 0.987 | 0.390 |
| 1000 | 0.955 | 0.223 | 0.965 | 0.224 | 0.960 | 0.224 | 0.976 | 0.255 | 0.986 | 0.258 |

$\kappa_1(0.5)=0.4 \quad \kappa_2(0.5)=0.4 \quad \theta=1$

$Se_1=0.76 \quad Sp_1=0.72 \quad Se_2=0.40 \quad Sp_2=0.943 \quad \varepsilon_1=0.0480 \quad \varepsilon_0=0.0206 \quad p=25\%$

|   | Wald *CI* | | Logarit. *CI* | | Fieller *CI* | | Bootstrap *CI* | | Bayesian *CI* | |
|---|---|---|---|---|---|---|---|---|---|---|
| *n* | *CP* | *AL* | *CP* | *AL* | *CP* | *AL* | *CP* | *AL* | *CP* | *AL* |
| 25 | 0.979 | 1.627 | 0.999 | 1.835 | 0.990 | 5.762 | 0.977 | 2.244 | 0.999 | 3.650 |
| 50 | 0.953 | 1.525 | 0.991 | 1.708 | 0.977 | 3.028 | 0.981 | 2.173 | 0.995 | 2.728 |
| 100 | 0.941 | 1.350 | 0.983 | 1.467 | 0.962 | 2.342 | 0.956 | 1.703 | 0.984 | 2.051 |
| 200 | 0.953 | 0.972 | 0.971 | 1.014 | 0.955 | 1.212 | 0.960 | 1.091 | 0.979 | 1.251 |
| 300 | 0.950 | 0.770 | 0.953 | 0.790 | 0.944 | 0.851 | 0.941 | 0.825 | 0.965 | 0.931 |
| 400 | 0.955 | 0.658 | 0.969 | 0.670 | 0.960 | 0.705 | 0.959 | 0.694 | 0.980 | 0.776 |
| 500 | 0.951 | 0.582 | 0.954 | 0.590 | 0.947 | 0.612 | 0.943 | 0.607 | 0.965 | 0.678 |
| 1000 | 0.952 | 0.403 | 0.955 | 0.406 | 0.951 | 0.413 | 0.950 | 0.410 | 0.972 | 0.458 |

$\kappa_1(0.9)=0.4 \quad \kappa_2(0.9)=0.4 \quad \theta=1$

$Se_1=0.943 \quad Sp_1=0.229 \quad Se_2=0.70 \quad Sp_2=0.70 \quad \varepsilon_1=0.0200 \quad \varepsilon_0=0.0343 \quad p=50\%$

|   | Wald *CI* | | Logarit. *CI* | | Fieller *CI* | | Bootstrap *CI* | | Bayesian *CI* | |
|---|---|---|---|---|---|---|---|---|---|---|
| *n* | *CP* | *AL* | *CP* | *AL* | *CP* | *AL* | *CP* | *AL* | *CP* | *AL* |
| 25 | 1 | 1.857 | 1 | 2.233 | 1 | 4.483 | 1 | 2.595 | 1 | 4.216 |
| 50 | 0.999 | 1.762 | 0.999 | 2.134 | 0.997 | 3.455 | 0.979 | 1.943 | 1 | 3.294 |
| 100 | 0.995 | 1.685 | 0.997 | 1.876 | 0.992 | 2.338 | 0.974 | 1.770 | 0.997 | 2.396 |
| 200 | 0.983 | 1.195 | 0.988 | 1.278 | 0.980 | 1.345 | 0.980 | 1.268 | 0.990 | 1.445 |
| 300 | 0.964 | 0.943 | 0.982 | 0.986 | 0.959 | 1.003 | 0.965 | 0.989 | 0.971 | 1.093 |
| 400 | 0.957 | 0.803 | 0.976 | 0.828 | 0.951 | 0.838 | 0.957 | 0.839 | 0.971 | 0.913 |
| 500 | 0.954 | 0.709 | 0.970 | 0.726 | 0.956 | 0.733 | 0.960 | 0.739 | 0.970 | 0.801 |
| 1000 | 0.956 | 0.491 | 0.964 | 0.496 | 0.956 | 0.499 | 0.959 | 0.505 | 0.969 | 0.545 |

Similar conclusions are obtained when the dependence factors take high values. Therefore, regarding the effect of the dependence factors on the *CIs*, in general terms they do not have a clear effect on the *CPs* of the *CIs*.

*4.3. CIs with a small sample*

The results of the simulation experiments have shown that the *CIs* may fail when the sample size is small $(n = 25-50)$. A classic solution to this problem is adding the correction 0.5 to each observed frequency, as is frequent in the analysis of $2 \times 2$ tables. To assess this procedure, the same simulation experiments as before were carried out for $n = \{25, 50, 100\}$ adding the value 0.5 to all of the observed frequencies $s_{ij}$ and $r_{ij}$. Table 7 shows some of the results obtained for the *CIs* for the ratio $\theta$. The results for the difference $\delta$ are not shown since, although this method improves the *CP* of the *CIs*, these intervals continue to fail when they failed without adding the correction. The results for $n = 100$ are not shown either, since these are very similar to those obtained without adding the correction.

Table 7. Coverage probabilities (*CPs*) and average lengths (*ALs*) of the *CIs* for $\theta$ with small samples.

| | $\kappa_1(0.9) = 0.2$  $\kappa_2(0.9) = 0.8$  $\theta = 0.25$ | | | | | | | | | |
|---|---|---|---|---|---|---|---|---|---|---|
| | $Se_1 = 0.28$  $Sp_1 = 0.92$  $Se_2 = 0.82$  $Sp_2 = 0.98$  $\varepsilon_1 = 0.0252$  $\varepsilon_0 = 0.00092$  $p = 10\%$ | | | | | | | | | |
| | Wald *CI* | | Logarit. *CI* | | Fieller *CI* | | Bootstrap *CI* | | Bayesian *CI* | |
| n | CP | AL | CP | AL | CP | AL | CP | AL | CP | AL |
| 25 | 0.999 | 1.808 | 0.008 | 1.960 | 0.653 | 3.014 | 0.145 | 2.150 | 0.783 | 3.531 |
| 50 | 0.940 | 1.287 | 0.262 | 1.464 | 0.768 | 1.710 | 0.556 | 1.440 | 0.768 | 1.813 |
| | $\kappa_1(0.5) = 0.4$  $\kappa_2(0.5) = 0.8$  $\theta = 0.5$ | | | | | | | | | |
| | $Se_1 = 0.76$  $Sp_1 = 0.72$  $Se_2 = 0.85$  $Sp_2 = 0.95$  $\varepsilon_1 = 0.0570$  $\varepsilon_0 = 0.0180$  $p = 25\%$ | | | | | | | | | |
| | Wald *CI* | | Logarit. *CI* | | Fieller *CI* | | Bootstrap *CI* | | Bayesian *CI* | |
| n | CP | AL | CP | AL | CP | AL | CP | AL | CP | AL |
| 25 | 1 | 1.458 | 0.961 | 1.659 | 0.984 | 2.332 | 0.940 | 1.897 | 1 | 3.118 |
| 50 | 0.992 | 0.836 | 0.960 | 0.913 | 0.982 | 0.932 | 0.962 | 0.869 | 0.997 | 1.141 |
| | $\kappa_1(0.9) = 0.6$  $\kappa_2(0.9) = 0.8$  $\theta = 0.75$ | | | | | | | | | |
| | $Se_1 = 0.62$  $Sp_1 = 0.98$  $Se_2 = 0.911$  $Sp_2 = 0.936$  $\varepsilon_1 = 0.0277$  $\varepsilon_0 = 0.0094$  $p = 5\%$ | | | | | | | | | |
| | Wald *CI* | | Logarit. *CI* | | Fieller *CI* | | Bootstrap *CI* | | Bayesian *CI* | |
| n | CP | AL | CP | AL | CP | AL | CP | AL | CP | AL |
| 25 | 1 | 1.812 | 1.000 | 2.073 | 1 | 3.554 | 1 | 2.425 | 1 | 4.053 |
| 50 | 1 | 1.593 | 1.000 | 1.789 | 1 | 2.564 | 0.999 | 2.067 | 1 | 2.682 |
| | $\kappa_1(0.9) = 0.4$  $\kappa_2(0.9) = 0.4$  $\theta = 1$ | | | | | | | | | |
| | $Se_1 = 0.943$  $Sp_1 = 0.229$  $Se_2 = 0.70$  $Sp_2 = 0.70$  $\varepsilon_1 = 0.0200$  $\varepsilon_0 = 0.0343$  $p = 50\%$ | | | | | | | | | |
| | Wald *CI* | | Logarit. *CI* | | Fieller *CI* | | Bootstrap *CI* | | Bayesian *CI* | |
| n | CP | AL | CP | AL | CP | AL | CP | AL | CP | AL |
| 25 | 1 | 1.896 | 1 | 2.140 | 1 | 4.727 | 1 | 2.571 | 1 | 4.234 |
| 50 | 1 | 1.798 | 1 | 1.991 | 1 | 3.211 | 1 | 2.418 | 1 | 3.242 |

As conclusions, in general terms, it holds that: a) the Wald *CI* for $\theta$ does not fail, its *CP* is 100% or very close to 100%, and its *AL* is lower than the rest of the intervals when these do not fail; b) the logarithmic, Fieller, Bootstrap and Bayesian *CIs* may continue to fail when $\theta = 0.25$. Consequently, when the sample size is small one must use the Wald *CI* for $\theta$ adding the value 0.5 to all of the observed frequencies.

*4.4. Rules of application*

The *CIs* for the difference and for the ratio of the two weighted kappa coefficients compare both parameters, and therefore we can decide which method is preferable to make this comparison. Once we have studied the coverage probabilities and the average lengths of the *CIs* for $\delta = \kappa_1(c) - \kappa_2(c)$ and for $\theta = \kappa_1(c)/\kappa_2(c)$, from the results obtained some general rules of application can be given for the *CIs* in terms of sample size. These rules are based on the failures and on the coverage probabilities, since the average lengths of the *CIs* for the difference and for the ratio cannot be compared as they are different intervals. In terms of sample size *n*:

a) If *n* is small $(n < 100)$, use the Wald *CI* for $\theta$ increasing the frequencies $s_{ij}$ and $r_{ij}$ in 0.5.

b) If $100 \leq n \leq 400$, use the Wald *CI* for the ratio $\theta$ without adding 0.5.

c) If $n \geq 500$, use any of the *CIs* (for the difference or for the ratio) proposed in Section 3.2 without adding 0.5.

In general terms, if the sample size is small, the Wald *CI* calculated adding 0.5 to each observed frequency does not fail. In this situation, its *AL* increases in relation to the Wald *CI* without adding 0.5, but its *CP* also increases meaning that the interval does not fail. When $100 \leq n \leq 400$ the *CI* that behaves best (fewest failures and its *CP* shows better fluctuations around 95%) is the Wald *CI* for the ratio $\theta$. When the sample size is very large $(n \geq 500)$, there is no important difference between the asymptotic behaviour of the proposed *CIs*, and therefore any one of them can be used. When the sample size is small, $(n \leq 50)$ the *CIs* may fail, especially when the difference between the two weighted kappa coefficients is not small.

## 5. Sample size

The determination of the sample size to compare parameters of two *BDTs* is a topic of interest. We then propose a method to calculate the sample size to estimate the ratio $\theta$ between two weighted kappa coefficients with a precision $\phi$ and a confidence $100(1-\alpha)\%$. This method is based on the Wald *CI* for $\theta$, which is, in general terms, the interval with the best asymptotic behaviour. Furthermore, this method requires a pilot sample (or another previous study) from which we calculate estimations of all of the parameters ($Se_i$, $Sp_i$, $\varepsilon_i$ and $p$, and consequently of $\kappa_i(c)$) and the Wald *CI* for $\theta$. If the pilot sample size is not small and the Wald *CI* for $\theta$ calculated from this sample contains the value 1, it makes no sense to determine the sample size necessary to estimate how much bigger one weighted kappa coefficient is than the other one, as the equality between both is not rejected. Nevertheless, if the pilot sample is small and the Wald *CI* (adding 0.5) contains the value 1, it may be useful to calculate the sample size to estimate the ratio $\theta$. In this situation, the Wald *CI* (adding 0.5) will be very wide (as the pilot sample is small) and may contain the value 1 even if $\kappa_1(c)$ and $\kappa_2(c)$ are different. Let us considerer that $\kappa_2(c) \geq \kappa_1(c)$ and therefore $\theta \leq 1$, and let $\phi$ be the precision set by the researcher. As it has been assumed that $\theta \leq 1$, then $\phi$ must be lower than one, and if we want to have a high level of precision then $\phi$ must be a small value.

On the other and, based on the asymptotic normality of $\hat{\theta} = \hat{\kappa}_1(c)/\hat{\kappa}_2(c)$ it is verified that $\hat{\theta} \in \theta \pm z_{1-\alpha/2}\sqrt{Var(\hat{\theta})}$, i.e. the probability of obtaining an estimator $\hat{\theta}$ is in this interval with a probability $100(1-\alpha)\%$. Setting a precision $\phi$, we can then calculate the sample size $n$ from

$$\phi = z_{1-\alpha/2}\sqrt{Var(\hat{\theta})}. \tag{21}$$

where

$$Var(\hat{\theta}) \approx \frac{\kappa_2^2(c)Var[\hat{\kappa}_1(c)] + \kappa_1^2(c)Var[\hat{\kappa}_2(c)] - 2\kappa_1(c)\kappa_2(c)Cov[\hat{\kappa}_1(c),\hat{\kappa}_2(c)]}{\kappa_2^4(c)}.$$

In the Appendix *B* of the supplementary material, we can see how this expression is obtained. This variance depends on the weighted kappa coefficients and on their respective variances and covariance. Furthermore, the variances $Var[\hat{\kappa}_i(c)]$ and the covariance $Cov[\hat{\kappa}_1(c),\hat{\kappa}_2(c)]$ (their expressions can be seen in the Appendix B of the supplementary material) depend, among other parameters, on the sample size *n*. Consequently, it is possible to use this relation to calculate the sample size to estimate the ratio $\theta$. Substituting in the equation of $Var(\hat{\theta})$ the variances and the covariance with its respective expressions, substituting the parameters with their estimators and clearing *n* in equation (21), it is obtained that

$$n = \frac{z_{1-\alpha/2}^2 \hat{\theta}^2}{\phi^2 \hat{p}^3 \hat{q}^3} \times$$

$$\left\{ \sum_{h=1}^{2} \left[ \frac{\hat{a}_{h1}^2 \hat{S}e_h(1-\hat{S}e_h)\hat{q} + \hat{a}_{h2}^2 \hat{S}p_h(1-\hat{S}p_h)\hat{p} + \hat{a}_{h3}^2 \hat{p}^2 \hat{q}^2}{\hat{Y}_h^2} \right] - \frac{2}{\hat{Y}_1 \hat{Y}_2} \left[ \hat{a}_{11}\hat{a}_{21}\hat{\varepsilon}_1 \hat{q} + \hat{a}_{12}\hat{a}_{22}\hat{\varepsilon}_0 \hat{p} + \hat{a}_{13}\hat{a}_{23}\hat{p}^2 \hat{q}^2 \right] \right\}, \tag{22}$$

where $\hat{a}_{h1} = \hat{p}\hat{q} - \hat{p}(\hat{q}-c)\hat{\kappa}_h(c)$, $\hat{a}_{h2} = \hat{a}_{h1} + (\hat{q}-c)\hat{\kappa}_h(c)$ and

$\hat{a}_{h3} = (1-2\hat{p})\hat{Y}_h - \left[(1-c-2\hat{p})\hat{Y}_h + \hat{S}p_h + c - 1\right]\hat{\kappa}_h(c)$. This method requires us to know $\hat{S}e_h$, $\hat{S}p_h$, $\hat{\varepsilon}_i$ and $\hat{p}$ (and therefore $\hat{\kappa}_h(c)$), for example obtained from a pilot sample or from previous studies. The procedure to calculate the sample size consists of the following steps:

1) Take pilot samples sized $n'$ (in general terms, $n' \geq 100$ to be able to calculate the Wald *CI* without adding 0.5 or use the Wald *CI* adding 0.5 to the frequencies if $n$ is small), and from this sample calculate $\hat{S}e_h$, $\hat{S}p_h$, $\hat{\varepsilon}_i$, $\hat{p}$ and $\hat{\kappa}_h(c)$, and a then calculate the Wald *CI* for $\theta$. If the Wald *CI* calculated has a precision $\phi$, i.e. if $\frac{\text{Upper limit} - \text{Lower limit}}{2} \leq \phi$, then with the pilot sample the precision has been reached and the process has finished ($\theta$ has been estimated with a precision $\phi$ to a confidence $100(1-\alpha)\%$); if this is not the case, go to the following step.

2) From the estimations obtained in Step 1, calculate the new sample size $n$ applying equation (22).

3) Take the sample of $n$ individuals ($n - n'$ is added to the pilot sample), and from the new sample we calculate $\hat{S}e_h$, $\hat{S}p_h$, $\hat{\varepsilon}_i$, $\hat{p}$, $\hat{\kappa}_h(c)$ and the Wald *CI* for $\theta$. If the Wald *CI* calculated has a precision $\phi$, then with the new sample the precision has been reached and the process has finished. If the Wald *CI* does not have the required precision, then this new sample is considered as a pilot sample and the process starts again at step 1. In this situation, the new sample has a size $n$ calculated in step 2, i.e. we add $n - n'$ individuals to the initial pilot sample (sized $n'$). Therefore, the process starts again at step 1 considering the new sample as the pilot sample and from this sample we calculate the values of the estimators and the Wald *CI*.

The method to calculate the sample size is an iterative method which depends on the pilot sample and which does not guarantee that $\theta$ will be estimated with the required precision. Each time that the previous process (steps 1-3) is repeated, we calculate (starting from an initial sample) the new sample size to estimate $\theta$, i.e. we calculate the number of individuals that must be added to the initial sample to obtain a new sample. Therefore, this process adjusts the size of the initial pilot sample, adding (in each iteration of the process: steps 1-3) the number of individuals necessary to obtain the right sample size to estimate $\theta$ with the precision required. The programme in *R* described in the Section 6 allows us to calculate the sample size to estimate $\theta$.

If the Wald *CI* for $\theta$ is higher than one, the *BDTs* can always be permuted and $\theta$ will then be lower than one. Another alternative consists of setting a value for a precision $\phi'$, in a similar way to the previous situation when $\theta \leq 1$, and then apply the equation (22) with $\phi = \hat{\theta}^2 \phi'$, where $\hat{\theta} = \hat{\kappa}_1(c)/\hat{\kappa}_2(c) \leq 1$. This is due to the fact that if $(L_\theta, U_\theta)$ is the Wald *CI* for $\theta = \kappa_1(c)/\kappa_2(c) \leq 1$ then the Wald *CI* for $\theta' = 1/\theta = \kappa_2(c)/\kappa_1(c)$ is $(L_\theta/\hat{\theta}^2, U_\theta/\hat{\theta}^2)$. It is easy to check that the calculated value of the sample size *n* is the same both if $\theta \leq 1$ (with precision $\phi$) and if $\theta > 1$ (with precision $\phi = \hat{\theta}^2 \phi'$).

Simulation experiments were carried out to study the effect that the pilot sample has on the calculation of the sample size. These experiments consisted of generating $N = 10,000$ random samples of multinomial distributions considering the same scenarios as those given in Tables 5 and 6. The equation of the sample size depends on the values of the estimators, which in turn depend on the pilot sample. Consequently, the pilot sample may have an effect on the sample size calculated. To study this effect, the simulation experiments consisted of the following steps:

1) Calculate the sample size $n$ from the values of the parameters set in the different scenarios considered. Therefore, equation (22) was applied using the values of the parameters (instead of their estimators).

2) Generate the $N$ multinomial random samples sized $n$ calculating the probabilities from equations (6) and (7), using the values of the previous parameters, and as $\varepsilon_i$ we considered low values (25%), intermediate values (50%) and high values (80%). From each one of the $N$ random samples, $\hat{Se}_h$, $\hat{Sp}_h$, $\hat{\varepsilon}_i$ and $\hat{p}$ (and therefore $\hat{\kappa}_h(c)$) were calculated, and then we calculated the sample size $n'_i$ applying equation (22).

3) For each scenario, the average sample size and the relative bias were calculated, i.e. $\bar{n} = \sum n'_i / N$ and $RB(n') = (\bar{n} - n)/n$.

Table 7. Effect of the pilot sample on the sample size.

| | $\kappa_1(0.1)=0.2$ $\kappa_2(0.1)=0.8$ $\theta=0.25$ | | | | | |
|---|---|---|---|---|---|---|
| | $Se_1=0.484$ $Sp_1=0.684$ $Se_2=0.852$ $Sp_2=0.911$ $p=50\%$ | | | | | |
| | $\varepsilon_1=0.0179$ $\varepsilon_0=0.0153$ | | $\varepsilon_1=0.0359$ $\varepsilon_0=0.0306$ | | $\varepsilon_1=0.0574$ $\varepsilon_0=0.0489$ | |
| | $\phi=0.05$ | $\phi=0.10$ | $\phi=0.05$ | $\phi=0.10$ | $\phi=0.05$ | $\phi=0.10$ |
| Sample size | 3170 | 793 | 3066 | 767 | 2942 | 736 |
| Average sample size | 3173 | 795 | 3068 | 769 | 2946 | 738 |
| Relative bias (%) | 0.095 | 0.252 | 0.065 | 0.261 | 0.136 | 0.272 |
| | $\kappa_1(0.9)=0.2$ $\kappa_2(0.9)=0.8$ $\theta=0.25$ | | | | | |
| | $Se_1=0.28$ $Sp_1=0.92$ $Se_2=0.82$ $Sp_2=0.98$ $p=10\%$ | | | | | |
| | $\varepsilon_1=0.0126$ $\varepsilon_0=0.0046$ | | $\varepsilon_1=0.0252$ $\varepsilon_0=0.0092$ | | $\varepsilon_1=0.0403$ $\varepsilon_0=0.0147$ | |
| | $\phi=0.05$ | $\phi=0.10$ | $\phi=0.05$ | $\phi=0.10$ | $\phi=0.05$ | $\phi=0.10$ |
| Sample size | 5104 | 1276 | 4947 | 1237 | 4758 | 1190 |
| Average sample size | 5113 | 1287 | 4948 | 1246 | 4759 | 1218 |
| Relative bias (%) | 0.18 | 0.83 | 0.02 | 0.73 | 0.02 | 2.35 |

Table 7 (Effect of the pilot sample) shows some of the results obtained. The relative biases are very small, which indicates that the equation of the calculation of the sample size provides robust values, and therefore the choice of the pilot sample does not have an important effect on the calculation of the sample size.

## 6. Programme *citwkc*

A programme has been written in *R* and called citwkc (Confidence Intervals for Two Weighted Kappa Coefficients) which allows us to calculate the *CIs* proposed in Section 3 and the sample size proposed in Section 5. The programme runs with the command

$$\text{citwkc}\left(s_{11}, s_{10}, s_{01}, s_{00}, r_{11}, r_{10}, r_{01}, r_{00}, cindex, preci = 0, conf = 0.95\right),$$

where *cindex* is the weighting index, *preci* is the precision that is needed to calculate the sample size and *conf* is the level of confidence (by default 95%). By default $preci = 0$, and the programme does not calculate the sample size, and only calculates it when $preci > 0$. In this situation $\left(preci > 0\right)$, the programme checks if it is necessary to calculate the sample size. The programme checks that the values of the frequencies and of the parameters are viable (e.g. that there are no negative values, frequencies with decimals, etc…), and also checks that it is possible to estimate all of the parameters and their variances-covariances. For the intervals obtained applying the Bootstrap method, 2,000 samples with replacement are generated, and for the Bayesian intervals 10,000 random samples are generated. The results obtained on running the programme are saved in file called "Results_citwkc.txt" in the same folders from where the programme is run. The program is available for free at URL

"https://www.ugr.es/~bioest/software/cmd.php?seccion=mdb".

## 7. Application

The results obtained have been applied to the study by Batwala et al (2010) on the diagnosis of malaria. Batwala et al have applied the *Expert Microscopy Test* and the *HRP2-Based Rapid Diagnostic Test* to a sample of 300 individuals using the *PCR* as the *GS*. The results of the study are shown in Table 8, where the $T_1$ models the result of the *Expert Microscopy Test*, $T_2$ models the result of the *HRP2-Based Rapid Diagnostic Test* and *D* models the result of the *PCR*. In this example, $\hat{Se}_1 = 46.07\%$, $\hat{Sp}_1 = 97.16\%$, $\hat{Se}_2 = 91.01\%$ and $\hat{Sp}_2 = 86.26\%$, and therefore $\widehat{rTPF}_{12} = 0.506$ and $\widehat{rFPF}_{12} = 0.207$. Applying the equation (5) it holds that $c' = 0.1902$. As $\widehat{rTPF}_{12} < 1$ and $\widehat{rFPF}_{12} < 1$, applying the rule c) given in Section 2, it holds that $\hat{\kappa}_1(c) > \hat{\kappa}_2(c)$ for $0 \leq c < 0.1902$ and that $\hat{\kappa}_1(c) < \hat{\kappa}_2(c)$ for $0.1902 < c \leq 1$. Table 8 also shows the values of $\hat{\kappa}_h(c)$, $\hat{\delta}$, $\hat{\theta}$ and the 95% *CIs* when $c = \{0.1, 0.1902, 0.2, ..., 0.8, 0.9\}$. The results were obtained running the programme "*citwkc*" with the command "$citwkc(41, 0, 40, 8, 5, 1, 24, 181, c)$" taking $c = \{0.1, 0.1902, 0.2, ..., 0.8, 0.9\}$. Applying the rules given in Section 4, as $n = 300 < 400$ then it is necessary to use the Wald *CI* for the ratio $\theta$.

For $c = \{0.1, 0.1902, 0.2, 0.3\}$, the Wald *CI* for $\theta$ contains the value 1, and therefore in these cases we do not reject the equality of the weighted kappa coefficients of the *Expert Microscopy Test* and of the *HRP2-Based Rapid Diagnostic Test*. Therefore, when the clinician considers that a false positive is 9, 4 or 2.33 times more important than a false negative, we do not reject the equality between the weighted kappa coefficients of the *Expert Microscopy Test* and of the *HRP2-Based Rapid Diagnostic Test* in the population studied. The rest of the intervals for $\theta$ also contain the value 1 and all of the *CIs* for the difference $\delta$ contain the value 0, and therefore the same conclusion is obtained.

For $c = \{0.4, 0.5, ..., 0.8, 0.9\}$, the Wald *CI* $\theta$ does not contain the value 1, and therefore in all of these cases we do not reject the equality of the weighted kappa coefficients of the *Expert Microscopy Test* and of the *HRP2-Based Rapid Diagnostic Test* in the population studied. Therefore, the clinician considers that a false negative is more important than a false positive (as happens in the situation in which the diagnostic tests are applied as screening tests), the weighted kappa coefficient of the *HRP2-Based Rapid Diagnostic Test* is significantly greater than the weighted kappa coefficient of the *Expert Microscopy Test* in the population studied. The same conclusion is obtained when the clinician considers that a false positive and a false negative have the same importance $(c = 0.5)$. If the clinician considers that a false positive is 1.5 times greater than a false negative (i.e. $c = 0.4$), then the same conclusion is obtained. The rest of the *CIs* for $\theta$ do not contain the value 1, nor do the three *CIs* for the difference $\delta$ contain the value 0, and therefore the same conclusions are obtained as with the Wald *CI* for $\theta$. For example, considering $c = 0.9$, interpreting the Wald *CI* for the ratio, it is concluded that in the population being studied the between the *HRP2-Based Rapid Diagnostic Test* and the *PCR* is, with a confidence of 95%, a value between 1.72 $(1/0.58 \approx 1.72)$ and 2.94 $(1/0.34 \approx 2.94)$ times greater than the agreement beyond chance between the *Expert Microscopy Test* and the *PCR*.

In order to illustrate the method to calculate the sample size presented in Section 5 we will consider that $c = 0.9$, and therefore that the two *BDTs* are applied as a screening test. In this situation, the 95% Wald *CI* for $\theta$ is $(0.34, 0.58)$, and the precision is 0.12. As an example, we will consider that the clinician wishes to estimate the ratio between the two weighted kappa coefficients with a precision $\phi = 0.10$. As with the sample of 300 individuals the desired precision $(\phi = 0.10 < 0.12)$ was not achieved, then using this sample as a pilot sample and running the programme *citwkc* with the command "citwkc$(41,0,40,8,5,1,24,181,0.9,0.1)$" it holds that $n = 435$. Therefore, to the sample pilot of 300 individuals we must add 135 more. Once the new sample has been taken, it is necessary to check that the precision $\phi = 0.10$ is verified.

Table 8. Study of Batwala et al and results.

| | | | $T_1 = 1$ | | $T_1 = 0$ | | |
| --- | --- | --- | --- | --- | --- | --- | --- |
| | | | $T_2 = 1$ | $T_2 = 0$ | $T_2 = 1$ | $T_2 = 0$ | Total |
| | $D = 1$ | | 41 | 0 | 40 | 8 | 89 |
| | $D = 0$ | | 5 | 1 | 24 | 181 | 211 |
| | Total | | 46 | 1 | 64 | 189 | 300 |

| | | | | 95% CIs for the difference $\delta = \kappa_1(c) - \kappa_2(c)$ | | |
| --- | --- | --- | --- | --- | --- | --- |
| $c$ | $\hat{\kappa}_1(c)$ | $\hat{\kappa}_2(c)$ | $\hat{\delta}$ | Wald | Bias corrected | Bayesian |
| 0.1 | 0.726 | 0.642 | 0.084 | -0.041 , 0.208 | -0.051 , 0.200 | -0.080 , 0.219 |
| 0.1902 | 0.659 | 0.659 | 0 | -0.125 , 0.125 | -0.130 , 0.124 | -0.155 , 0.136 |
| 0.2 | 0.653 | 0.661 | -0.008 | -0.133 , 0.116 | -0.137 , 0.117 | -0.162 , 0.128 |
| 0.3 | 0.593 | 0.681 | -0.088 | -0.213 , 0.037 | -0.214 , 0.039 | -0.233 , 0.051 |
| 0.4 | 0.543 | 0.701 | -0.158 | -0.283 , -0.034 | -0.284 , -0.032 | -0.298 , -0.018 |
| 0.5 | 0.501 | 0.723 | -0.222 | -0.345 , -0.100 | -0.347 , -0.100 | -0.357 , -0.081 |
| 0.6 | 0.464 | 0.747 | -0.283 | -0.402 , -0.163 | -0.402 , -0.163 | -0.411 , -0.140 |
| 0.7 | 0.433 | 0.772 | -0.339 | -0.455 , -0.223 | -0.454 , -0.222 | -0.461 , -0.195 |
| 0.8 | 0.406 | 0.799 | -0.393 | -0.506 , -0.280 | -0.504 , -0.276 | -0.511 , -0.247 |
| 0.9 | 0.382 | 0.827 | -0.445 | -0.557 , -0.333 | -0.557 , -0.329 | -0.561 , -0.296 |

| | | | | 95% CIs for the ratio $\theta = \kappa_1(c)/\kappa_2(c)$ | | | | |
| --- | --- | --- | --- | --- | --- | --- | --- | --- |
| $c$ | $\hat{\kappa}_1(c)$ | $\hat{\kappa}_2(c)$ | $\hat{\theta}$ | Wald | Logarithmic | Fieller | Bias-corrected | Bayesian |
| 0.1 | 0.726 | 0.642 | 1.131 | 0.925 , 1.335 | 0.943 , 1.355 | 0.940 , 1.357 | 0.926 , 1.344 | 0.883 , 1.393 |
| 0.1902 | 0.659 | 0.659 | 1 | 0.811 , 1.189 | 0.828 , 1.208 | 0.823 , 1.206 | 0.817 , 1.204 | 0.776 , 1.234 |
| 0.2 | 0.653 | 0.661 | 0.988 | 0.800 , 1.174 | 0.817 , 1.194 | 0.812 , 1.192 | 0.808 , 1.192 | 0.766 , 1.219 |
| 0.3 | 0.593 | 0.681 | 0.871 | 0.695 , 1.046 | 0.711 , 1.065 | 0.704 , 1.059 | 0.701 , 1.065 | 0.673 , 1.083 |
| 0.4 | 0.543 | 0.701 | 0.775 | 0.609 , 0.939 | 0.625 , 0.958 | 0.615 , 0.948 | 0.615 , 0.952 | 0.593 , 0.971 |
| 0.5 | 0.501 | 0.723 | 0.693 | 0.537 , 0.847 | 0.553 , 0.866 | 0.541 , 0.854 | 0.541 , 0.857 | 0.525 , 0.877 |
| 0.6 | 0.464 | 0.747 | 0.621 | 0.476 , 0.768 | 0.492 , 0.786 | 0.479 , 0.772 | 0.481 , 0.776 | 0.468 , 0.799 |
| 0.7 | 0.433 | 0.772 | 0.561 | 0.425 , 0.698 | 0.440 , 0.716 | 0.426 , 0.701 | 0.430 , 0.707 | 0.418 , 0.727 |
| 0.8 | 0.406 | 0.799 | 0.508 | 0.380 , 0.637 | 0.395 , 0.654 | 0.381 , 0.639 | 0.384 , 0.644 | 0.375 , 0.667 |
| 0.9 | 0.382 | 0.827 | 0.462 | 0.341, 0.582 | 0.356 , 0.599 | 0.342 , 0.584 | 0.347 , 0.594 | 0.339 , 0.611 |

## 8. Discussion

The weighted kappa coefficient of a *BDT* is a measure of the beyond-chance agreement between the *BDT* and the *GS*, and depends on the sensitivity and specificity of the *BDT*, on the disease prevalence and on the weighting index. The weighted kappa coefficient is a parameter that is used to assess and compare the performance of *BDTs*. In this article, we have studied the comparison of the weighted kappa coefficients of two *BDTs* through confidence intervals when the sample design is paired.

Three intervals have been studied for the difference of the two weighted kappa coefficients and five more intervals for the ratio of the two parameters. All the intervals studied are asymptotic and simulation experiments have been carried out to study their coverage probabilities and average lengths subject to different scenarios and for different sample sizes. Based on the results of the simulation experiments, some general rules of application have been given. When the sample size is moderate $(n=100)$ or large $(n=200-400)$ it is preferable to compare the two weighted kappa coefficients through an interval for the ratio, and when the sample size is very large $(n \geq 500)$ the two weighted kappa coefficients can be compared through the difference or the ratio. When the sample size is small $(n \leq 50)$, the interval with the best behaviour is the Wald *CI* for the ratio $\theta$ adding 0.5 to all of the observed frequencies. Adding 0.5 to all of the frequencies does not improve the behaviour of the intervals for the difference $\delta$, since these continue to fail when they failed without adding the value 0.5. This question may be due to the fact that the ratio $\hat{\theta}$ converges more quickly to the normal distribution than the difference $\hat{\delta}$. In the simulation experiments, the asymptotic behaviour of the Bayesian *CIs* has been studied using the *Beta*$(1,1)$ distribution as prior distribution for all of the parameters. The choice of the values of the hyperparameters of the *Beta* distribution will depend on the previous information that the researcher has. If the researcher has some information and wants this information to have some weight in the data, then it is possible to use higher values of $\alpha$ and $\beta$, i.e. considering a *Beta*$(\alpha,\beta)$ distribution with $\alpha,\beta >1$. The increase in $\alpha$ and $\beta$ adds information and decreases the variance and, therefore, there is less uncertainty about the parameter. If the researcher does not want this information to have a great weight in the posteriori distribution, then the researcher chooses moderate values of $\alpha$ and $\beta$ which are consistent with the information available, i.e. the average should be compatible with that information. To assess the effect that the *Beta* distribution has on the asymptotic behaviour of the Bayesian interval, we have carried out simulations (in a similar way to those carried out in Section 4) using as prior the distributions $Beta(5,5)$ and $Beta(25,25)$ for the Bayesian interval for $\theta = \kappa_1(c)/\kappa_2(c)$. These two distributions have the same average as the $Beta(1,1)$ distribution but different variances. The first distribution has a moderate weight in the subsequent distribution, the second has an important weight and the third one has a very important weight. In general terms, the results obtained with the distribution $Beta(5,5)$ are very similar to those obtained with the $Beta(1,1)$ distribution. Regarding the $Beta(25,25)$ distribution, there is no important difference in relation to the *CPs* obtained with the $Beta(1,1)$, although for $\theta = \{0.25, 0.50\}$ the *AL* is slightly lower with the $Beta(25,25)$, and when $\theta = \{0.75,1\}$ the *AL* is slightly higher with the $Beta(25,25)$. In general terms, when the Bayesian interval fails using the $Beta(1,1)$ distribution then it also fails using the $Beta(5,5)$ and the $Beta(25,25)$. Furthermore, the Bayesian *CI* for $\theta = \kappa_1(c)/\kappa_2(c)$ with the $Beta(5,5)$ and $Beta(25,25)$, respectively, does not display a better *CP* than the Wald *CI* (when it does not fail), and therefore the Bayesian *CI* does not improve the asymptotic behaviour of the Wald *CI*.

The application of the *CIs* requires the marginal frequencies *s* and *r* to be higher than zero. If the marginal frequency *s* (or *r*) is equal to zero, then it is not possible to estimate the weighted kappa coefficient of each *BDT*. Moreover, if a marginal frequency $s_{ij} + r_{ij}$ is equal to zero, then it is possible to calculate all of the *CIs* proposed; but not if two of these marginal frequencies are equal to zero. In this last situation, one of the weighted kappa coefficients (or both) is equal to zero, and the variance and the covariance are also equal to zero. If $s_{10} + r_{10} = s_{01} + r_{01} = 0$ then $\hat{\kappa}_1(c) = \hat{\kappa}_2(c)$ and $\hat{Var}\left(\hat{\kappa}_1(c)\right) = \hat{Var}\left(\hat{\kappa}_2(c)\right) = Cov\left(\hat{\kappa}_1(c), \hat{\kappa}_2(c)\right)$, and the frequentist intervals cannot be calculated. A solution to this problem is to add 0.5 to each observed frequency.

In this article, we have also proposed a method to calculate the sample size to estimate the ratio between the two weighted kappa coefficients with a determined precision and confidence. This method, based on the Wald *CI* for the ratio, is an iterative method, which starting from a pilot sample adds individuals to the sample until the *CI* has the set precision. From the initial sample we estimate a vector of parameters and in the second stage we calculate the sample size. Furthermore, the simulation experiments carried out to study the robustness of the method to calculate the sample size have shown that the method has practical validity and the choice of the pilot sample has very little effect on this method.

When the two diagnostic tests are continuous, for each cut off point of each estimated ROC curve there will be a value of $\hat{Se}_h$ and of $\widehat{FPF}_h$ (and therefore of $\hat{Sp}_h = 1 - \widehat{FPF}_h$), with $h = 1, 2$. Once the clinician has set the value of the weighting index, $\hat{\kappa}_1(c)$ and $\hat{\kappa}_2(c)$ are calculated and therefore the confidence intervals studied in Section 3 can be applied.

**Supplementary material: Appendices A, B and C**

Appendices *A*, *B* and *C* are available as supplementary material of the manuscript in the URL: https://www.ugr.es/~bioest/software/cmd.php?seccion=mdb.

**Acknowledgements**

This research was supported by the Spanish Ministry of Economy, Grant Number MTM2016-76938-P. We thank the referee, the Associate Editor and the Editor (M.

Isabel Fraga Alves) of REVSTAT Statistical Journal for their helpful comments that improved the quality of the paper.

# Supplementary material

## Appendix A

From now onwards, we are going to suppose that $0 < Se_h < 1$, $0 < Sp_h < 1$, $0 < p < 1$ and $q = 1 - p$. Performing algebraic operations it is verified that

$$\kappa_1(c) - \kappa_2(c) = \frac{pq}{D_1 D_2} v \qquad (1)$$

where $D_h = p(1-Q_h)c + qQ_h(1-c)$ is the denominator of $\kappa_h(c)$, with $h = 1, 2$, and

$$v = q\Delta_1 - c(\Delta_1 - p\Delta_2) \qquad (2)$$

where $\Delta_1 = Se_1(1-Sp_2) - Se_2(1-Sp_1)$ and $\Delta_2 = Y_1 - Y_2 = Se_1 - Se_2 + Sp_1 - Sp_2$. Then $\kappa_1(c) > \kappa_2(c)$ if $v > 0$, since $D_h > 0$. Solving equation $\kappa_1(c) - \kappa_2(c) = 0$ in $c$ it holds that

$$c' = c = \frac{q\Delta_1}{\Delta_1 - p\Delta_2}, \qquad (3)$$

being $c'$ a real value. From now onwards, the rules so that $\kappa_1(c) > \kappa_2(c)$, $\kappa_2(c) > \kappa_1(c)$ and $\kappa_1(c) = \kappa_2(c)$, considering that $i = 1$ and $j = 2$ (the demonstrations for $i = 2$ and $j = 1$ are analogous).

a) If $rTPF_{12} \geq 1$ and $rFPF_{12} < 1$, or $rTPF_{12} > 1$ and $rFPF_{12} \leq 1$, then $\kappa_1(c) > \kappa_2(c)$ for $0 \leq c \leq 1$.

Let us suppose in the first place that $rTPF_{12} = 1$ and that $rFPF_{12} < 1$, then $Se_1 = Se_2 = Se$ and $Sp_1 > Sp_2$. Substituting in equation (2) it holds that $v = (Sp_1 - Sp_2)[cp + (q-c)Se]$. Here $v > 0$ if $cp + (q-c)Se > 0$, since $Sp_1 > Sp_2$. If $c=0$ or $c=1$, then $cp+(q-c)Se>0$ since $qSe>0$, and $p(1-Se)>0$ is verified; and as $Sp_1>Sp_2$, then $v>0$ and $\kappa_1(c)>\kappa_2(c)$. Let us suppose that $0<c<1$ and $p\geq Se$, then $cp+(q-c)Se=c(p-Se)+qSe>0$, and it is verified that $v>0$ and $\kappa_1(c)>\kappa_2(c)$. If $p<Se$, then $cp+(q-c)Se=(1-c)(Se-p)+(1-Se)p>0$, since $(1-c)(Se-p)>0$ and $(1-Se)p>0$. Therefore, $v>0$ and $\kappa_1(c)>\kappa_2(c)$.

Let us now suppose that $rTPF_{12}>1$ and that $rFPF_{12}<1$, then $Se_1>Se_2$ and $Sp_1>Sp_2$. It is easy to check that when $c=0$ or $c=1$ it is verified that $v>0$ and, therefore, $\kappa_1(c)>\kappa_2(c)$. Moreover, as $rTPF_{12}>1$ and $rFPF_{12}<1$ then dividing both parameters $(rTPF_{12}/rFPF_{12}>1)$ it holds that $\dfrac{rTPF_{12}}{rFPF_{12}}=\dfrac{Se_1(1-Sp_2)}{Se_2(1-Sp_1)}>1$, verifying that $\Delta_1=Se_1(1-Sp_2)-Se_2(1-Sp_1)>0$. As $Se_1>Se_2$ and $Sp_1>Sp_2$ then $\Delta_2=Se_1-Se_2+Sp_1-Sp_2>0$. Furthermore, as it is verified that $Se_1>Se_2$ then $1-Se_1<1-Se_2$, and $0<\dfrac{1-Se_1}{1-Se_2}<1$. Moreover, as $\dfrac{Sp_1}{Sp_2}>1$ then $\dfrac{Sp_1}{Sp_2}-\dfrac{1-Se_1}{1-Se_2}=\dfrac{\Delta_3}{Sp_2(1-Se_2)}>0$, when $\Delta_3=(1-Se_2)Sp_1-(1-Se_1)Sp_2>0$. It is easy to check that $\Delta_1=\Delta_2-\Delta_3$, so that $\Delta_2>\Delta_1$. Equation (2) can be written as

$$v=(q-c)\Delta_1+cp\Delta_2. \qquad (4)$$

Let us suppose that $0<c<1$, then if $q\geq c$ it is verified that $v>0$ and $\kappa_1(c)>\kappa_2(c)$. Let us now suppose that $q<c$, then $q-c<0$. Equation (4) can be written as

$$v=-(c-q)\Delta_1+cp\Delta_2$$

being $c-q>0$. Let us suppose that

$$v<0\Rightarrow-(c-q)\Delta_1+cp\Delta_2<0,$$

so that

$$-(c-q)\Delta_1 < -cp\Delta_2 \Rightarrow (c-q)\Delta_1 > cp\Delta_2 \Rightarrow c-q > cp\frac{\Delta_2}{\Delta_1}.$$

As $\Delta_2 > \Delta_1$ then $\frac{\Delta_2}{\Delta_1} > 1$, so that

$$c-q > cp\frac{\Delta_2}{\Delta_1} > cp > 0,$$

from where we obtain

$$c-q-cp > 0. \qquad (5)$$

Performing algebraic operations

$$c-q-cp = q(c-1)$$

As $0 < c < 1$, $1-c > 0$ and $c-1 < 0$, then $q(c-1) < 0$, which is contradictory with expression (5). Therefore, if $q < c$ then $v > 0$ and $\kappa_1(c) > \kappa_2(c)$.

The demonstrations for $rTPF_{12} > 1$ and $rFPF_{12} \leq 1$ are performed following a similar process to the previous one.

b). If $rTPF_{12} > 1$ and $rFPF_{12} > 1$, then:

b.1) $\kappa_1(c) > \kappa_2(c)$ if $0 < c' < c \leq 1$

b.2) $\kappa_1(c) < \kappa_2(c)$ if $0 \leq c < c' < 1$

b.3) $\kappa_1(c) = \kappa_2(c)$ if $c = c'$, with $0 < c' < 1$

b.4) $\kappa_1(c) > \kappa_2(c)$ for $0 \leq c \leq 1$ if $c' < 0$ (or $c' > 1$) and $rTPF_{12} > rFPF_{12} > 1$

b.5) $\kappa_1(c) < \kappa_2(c)$ for $0 \leq c \leq 1$ if $c' < 0$ (or $c' > 1$) and $rFPF_{12} > rTPF_{12} > 1$

Firstly, we are going to demonstrate that $c'$ cannot be equal to 0 or to 1. As $rTPF > 1$ and $rFPF > 1$, then it is verified that $Se_1 > Se_2$ and $Sp_1 < Sp_2$. If $c' = 0$ then $\Delta_1 = 0$, and it is verified that

$$\frac{Se_1}{Se_2} \times \frac{1-Sp_2}{1-Sp_1} = 1,$$

which is incompatible with $rTPF > 1$ and $rFPF > 1$, since as $\frac{Se_1}{Se_2} > 1$ and $0 < \frac{1-Sp_2}{1-Sp_1} < 1$ then it is verified that $\frac{Se_1}{Se_2} \times \frac{1-Sp_2}{1-Sp_1} \neq 1$. Therefore $c'$ cannot be equal to 0 if $rTPF > 1$ and $rFPF > 1$. If $c' = 1$ then $\Delta_1 - \Delta_2 = Sp_2(1-Se_1) - Sp_1(1-Se_2) = 0$, and it is verified that

$$\frac{Sp_2}{Sp_1} \times \frac{1-Se_1}{1-Se_2} = 1,$$

which is incompatible with $rTPF > 1$ and $rFPF > 1$, since as $\frac{Sp_2}{Sp_1} > 1$ and $0 < \frac{1-Se_1}{1-Se_2} < 1$ then it is verified that $\frac{Sp_2}{Sp_1} \times \frac{1-Se_1}{1-Se_2} \neq 1$. Therefore, $c'$ cannot be equal to 1 if $rTPF > 1$ and $rFPF > 1$.

Let us consider that $0 < c' < 1$, then we must verify one of the two following: 1) $0 < q\Delta_1 < \Delta_1 - p\Delta_2$, or 2) $\Delta_1 - p\Delta_2 < q\Delta_1 < 0$. Condition 1 implies that $\Delta_1 > 0$ and $\Delta_1 > p\Delta_2$, and Condition 2 implies that $\Delta_1 < 0$ and $\Delta_1 < p\Delta_2$.

Moreover, as $Se_1 > Se_2$ and $Sp_1 < Sp_2$ (which implies that $1-Sp_1 > 1-Sp_2$) then $Q_1 > Q_2$. Furthermore, if $c = c'$ then performing algebraic operations, each weighted kappa coefficient is expressed as

$$\kappa_h(c') = \frac{Y_h}{\tau_h},$$

when $\tau_h = \frac{\Delta_1 - Q_h \Delta_2}{\Delta_1 - p\Delta_2}$, with $h = 1, 2$. As $Q_1 > Q_2$, then $\tau_2 - \tau_1 > 0$ if $\Delta_2 > 0$, and $\tau_2 - \tau_1 < 0$ if $\Delta_2 < 0$. If $\Delta_2 > 0$, then

$$\tau_2 - \tau_1 = \frac{\Delta_2 (Q_1 - Q_2)}{\Delta_1 - p\Delta_2} > 0 \Rightarrow \Delta_1 - p\Delta_2 > 0 \Rightarrow \Delta_1 > p\Delta_2 > 0.$$

If $\Delta_2 < 0$, then

$$\tau_2 - \tau_1 = \frac{\Delta_2 (Q_1 - Q_2)}{\Delta_1 - p\Delta_2} < 0 \Rightarrow \Delta_1 - p\Delta_2 > 0 \Rightarrow \Delta_1 > p\Delta_2.$$

Therefore, whether $\Delta_2 > 0$ or $\Delta_2 < 0$, it is always verified that $\Delta_1 > p\Delta_2$. This condition is only compatible with Condition 1 obtained by the fact that $0 < c' < 1$, i.e. $0 < q\Delta_1 < \Delta_1 - p\Delta_2$. Therefore, it is always verified that $\Delta_1 > 0$ and $\Delta_1 > p\Delta_2$.

Moreover, from equation (3) it holds that $q\Delta_1 = c'(\Delta_1 - p\Delta_2)$, so that substituting this expression in equation (2) it holds that

$$v = (\Delta_1 - p\Delta_2)(c' - c). \tag{6}$$

As $\Delta_1 > p\Delta_2$ then $\Delta_1 - p\Delta_2 > 0$. Based on equation (6), if $0 \leq c < c' < 1$ then $v > 0$ and $\kappa_1(c) > \kappa_2(c)$. If $0 < c' < c \leq 1$ then $v < 0$ and $\kappa_1(c) < \kappa_2(c)$, and if $c = c'$ (with $0 < c' < 1$) then $v = 0$ and $\kappa_1(c) = \kappa_2(c)$.

If $c' < 0$ then one of the following two conditions must be verified: 1) $0 < q\Delta_1 < \Delta_1 < p\Delta_2 < \Delta_2$, or 2) $\Delta_2 < p\Delta_2 < \Delta_1 < q\Delta_1 < 0$. Condition 1 implies that $\Delta_1 > 0$ and therefore $Se_1(1 - Sp_2) > Se_2(1 - Sp_1)$, and from this inequality it holds that

$$\frac{Se_1}{Se_2} > \frac{1 - Sp_1}{1 - Sp_2} > 1 \Rightarrow rTPF_{12} > rFPF_{12} > 1.$$

As $q\Delta_1 > 0$ and $\Delta_1 - p\Delta_2 < 0$, then applying equation (2) it holds that $v > 0$ and therefore $\kappa_1(c) > \kappa_2(c)$. Condition 2 implies that $\Delta_1 < 0$ and therefore $Se_1(1 - Sp_2) < Se_2(1 - Sp_1)$, and it holds that

$$\frac{1 - Sp_1}{1 - Sp_2} > \frac{Se_1}{Se_2} > 1 \Rightarrow rFPF_{12} > rTPF_{12} > 1.$$

As $q\Delta_1 < 0$ and $\Delta_1 - p\Delta_2 > 0$, applying equation (2) again it holds that $v < 0$ and therefore $\kappa_1(c) < \kappa_2(c)$. If $c' > 1$, the demonstrations are similar to those of $c' < 0$.

c) If $rTPF_{12} < 1$ and $rFPF_{12} < 1$, then $rTPF_{21} > 1$ and $rFPF_{21} > 1$, and the demonstrations are analogous to case b).

**Appendix *B***

Bloch (1997) has deduced the expressions of the variances of $\hat{\kappa}_1(c)$ and $\hat{\kappa}_2(c)$ and of the covariance between them. We then obtain equivalent expressions and we also deduce the variance of the ratio of the two weighted kappa coefficients, an expression which is necessary to apply the method to calculate the sample size explained in Section 5. Let $\boldsymbol{\omega} = (Se_1, Sp_1, Se_2, Sp_2, p)^T$ be the vector of parameters, where $Se_1 = \frac{p_{10} + p_{11}}{p}$, $Sp_1 = \frac{q_{00} + q_{01}}{q}$, $Se_2 = \frac{p_{01} + p_{11}}{p}$ and $Sp_2 = \frac{q_{00} + q_{10}}{q}$, with $q = 1 - p$. Applying the delta method, the matrix of the asymptotic variances-covariances of $\hat{\boldsymbol{\omega}}$ is

$$\Sigma_{\hat{\boldsymbol{\omega}}} = \left(\frac{\partial \boldsymbol{\omega}}{\partial \boldsymbol{\pi}}\right) \Sigma_{\hat{\boldsymbol{\pi}}} \left(\frac{\partial \boldsymbol{\omega}}{\partial \boldsymbol{\pi}}\right)^T.$$

Performing the algebraic operations it is obtained that

$$Var(\hat{Se}_1) = \frac{(p_{11}+p_{10})(p_{01}+p_{00})}{np^3} = \frac{Se_1(1-Se_1)}{np},$$

$$Var(\hat{Se}_2) = \frac{(p_{11}+p_{01})(p_{10}+p_{00})}{np^3} = \frac{Se_2(1-Se_2)}{np},$$

$$Var(\hat{Sp}_1) = \frac{(q_{11}+q_{10})(q_{01}+q_{00})}{nq^3} = \frac{Sp_1(1-Sp_1)}{nq},$$

$$Var(\hat{Sp}_2) = \frac{(q_{11}+q_{01})(q_{10}+q_{00})}{nq^3} = \frac{Sp_2(1-Sp_2)}{nq}, \quad Var(\hat{p}) = \frac{pq}{n},$$

$$Cov[\hat{Se}_1, \hat{Se}_2] = \frac{p_{11}p_{00} - p_{10}p_{01}}{np^3} = \frac{\varepsilon_1}{np}, \quad Cov[\hat{Sp}_1, \hat{Sp}_2] = \frac{q_{11}q_{00} - q_{10}q_{01}}{nq^3} = \frac{\varepsilon_0}{nq}$$

and

$$Cov(\hat{Se}_h, \hat{Sp}_h) = Cov(\hat{Se}_h, \hat{p}) = Cov(\hat{Sp}_h, \hat{p}) = 0, \text{ with } h=1,2.$$

The estimators of the variances-covariances are obtained substituting each parameter with its corresponding estimator, where $\hat{Se}_1 = \frac{s_{11}+s_{10}}{s}$, $\hat{Se}_2 = \frac{s_{11}+s_{01}}{s}$, $\hat{Sp}_1 = \frac{r_{01}+r_{00}}{r}$,

$\hat{Sp}_2 = \frac{r_{10}+r_{00}}{r}$, $\hat{p} = \frac{s}{n}$, $\hat{q} = \frac{r}{n}$, $\hat{\varepsilon}_1 = \frac{\hat{p}_{11}}{\hat{p}} - \hat{Se}_1\hat{Se}_2 = \frac{s_{11}s_{00} - s_{10}s_{01}}{s^2}$ and

$\hat{\varepsilon}_0 = \frac{\hat{q}_{00}}{\hat{q}} - \hat{Sp}_1\hat{Sp}_2 = \frac{r_{11}r_{00} - r_{10}r_{01}}{r^2}$. Applying the delta method, the variance of $\hat{\kappa}_h(c)$ is

$$Var[\hat{\kappa}_h(c)] \approx \left[\frac{\partial \kappa_h(c)}{\partial Se_h}\right]^2 Var(\hat{Se}_h) + \left[\frac{\partial \kappa_h(c)}{\partial Sp_h}\right]^2 Var(\hat{Sp}_h) + \left[\frac{\partial \kappa_h(c)}{\partial p}\right]^2 Var(\hat{p}).$$

In this expression the covariances are zero. Performing the algebraic operations, it is obtained that

$$Var[\hat{\kappa}_h(c)] \approx \left[\frac{\kappa_h(c)}{pqY_h}\right]^2 \times \left[\{a_{h1}^2 Var(\hat{Se}_h) + a_{h2}^2 Var(\hat{Sp}_h) + a_{h3}^2 Var(\hat{p})\}\right]$$

with $h=1,2$, and where

$$a_{h1} = pq - p(q-c)\kappa_h(c),$$

$$a_{h2} = a_{h1} + (q-c)\kappa_h(c)$$

and

$$a_{h3} = (1-2p)Y_h - \left[(1-c-2p)Y_h + Sp_h + c - 1\right]\kappa_h(c).$$

The expression of $\hat{Var}\left[\hat{\kappa}_h(c)\right]$ is obtained substituting in the previous expressions each parameter with its estimator. Regarding the covariance between $\hat{\kappa}_1(c)$ and $\hat{\kappa}_2(c)$, applying the delta method again it is obtained that

$$Cov\left[\hat{\kappa}_1(c),\hat{\kappa}_2(c)\right] \approx \frac{\partial \kappa_1(c)}{\partial Se_1}\frac{\partial \kappa_2(c)}{\partial Se_2}Cov\left[\hat{Se}_1,\hat{Se}_2\right] + \frac{\partial \kappa_1(c)}{\partial Sp_1}\frac{\partial \kappa_2(c)}{\partial Sp_2}Cov\left[\hat{Sp}_1,\hat{Sp}_2\right] + \frac{\partial \kappa_1(c)}{\partial p}\frac{\partial \kappa_2(c)}{\partial p}Var(\hat{p}).$$

In this expression, the rest of the covariances are equal to zero. Performing the algebraic operations it is obtained that

$$Cov\left[\hat{\kappa}_1(c),\hat{\kappa}_2(c)\right] \approx$$
$$\frac{\kappa_1(c)\kappa_2(c)}{p^2q^2Y_1Y_2} \times \left[a_{11}a_{21}Cov\left(\hat{Se}_1,\hat{Se}_2\right) + a_{12}a_{22}Cov\left(\hat{Sp}_1,\hat{Sp}_2\right) + a_{13}a_{23}\hat{Var}(\hat{p})\right].$$

The expression of $\hat{Cov}\left[\hat{\kappa}_1(c),\hat{\kappa}_2(c)\right]$ is obtained substituting in this equation each parameter with its estimator.

Regarding the ration of the two weighted kappa coefficients, the variance of $\theta$ is easily calculated applying the delta method again, i.e.

$$Var(\hat{\theta}) \approx \sum_{h=1}^{2}\left(\frac{\partial \theta}{\partial \kappa_h(c)}\right)^2 Var\left[\hat{\kappa}_h(c)\right] + 2\frac{\partial \theta}{\partial \kappa_1(c)}\frac{\partial \theta}{\partial \kappa_2(c)}Cov\left[\hat{\kappa}_1(c),\hat{\kappa}_2(c)\right].$$

Performing the algebraic operations,

$$Var(\hat{\theta}) \approx \frac{\kappa_2^2(c)Var\left[\hat{\kappa}_1(c)\right] + \kappa_1^2(c)Var\left[\hat{\kappa}_2(c)\right] - 2\kappa_1(c)\kappa_2(c)Cov\left[\hat{\kappa}_1(c),\hat{\kappa}_2(c)\right]}{\kappa_2^4(c)}, \quad (7)$$

and substituting in this equation each parameter with its estimator, we obtain the expression of $\hat{Var}(\hat{\theta})$. The expression of variance of $\hat{Var}\left[\ln(\hat{\theta})\right]$ is calculated in a similar way to in the previous case, but considering $\ln(\theta)$ instead of $\theta$.

**Appendix *C***

The selection of the *CI* with the best asymptotic behaviour, both for the difference $\delta$ and for the ratio $\theta$, was made taking the following steps: 1) Choose the *CIs* with the least failures ($CP > 93\%$), 2) Choose the *CIs* that are the most accurate i.e. those with the lowest *AL*. The first step in this method establishes that the *CI* does not fail when $CP > 93\%$. In the simulation experiments the *CIs* were calculated to a 95% confidence i.e. $\gamma = 1 - \alpha = 0.95$ is the nominal confidence and $\alpha = 5\%$ is the nominal error. Then $\Delta\alpha = \alpha - \alpha^* = \gamma^* - \gamma$, where $\gamma^*$ is the *CP* calculated and $\alpha^*$ is the type I error.

Moreover, the hypothesis test to check the equality of the two weighted kappa coefficients is $H_0 : \kappa_1(c) = \kappa_2(c)$ vs $H_1 : \kappa_1(c) \neq \kappa_2(c)$. Based on the difference of both parameters, this hypothesis test is equivalent to test $H_0 : \delta = 0$ vs $H_1 : \delta \neq 0$. This test can be solved through different methods. Applying Bloch's method (1997), the test statistic is given by equation (equation (10) of the manuscript). The statistics for the bootstrap method and for the Bayesian method are obtained computationally.

In step 1 of the method, a *CI* has a failure if $CP \leq 93\%$, i.e. if $\Delta\alpha \leq -2$. In this situation, the type I error of the corresponding hypothesis test is $\geq 7\%$, and therefore it is a very liberal hypothesis test and it can give false significances. The criteria of 93% has been used by other authors (Price and Bonett, 2004; Martín-Andrés and Álvarez-Hernández, 2014a, 2014b; Montero-Alonso and Roldán-Nofuentes, 2018). If $\Delta\alpha > 2\%$, i.e. $CP > 97\%$, then the hypothesis test is very conservative (its type I error is very small, <3% ), but it does not give false significances. Consequently, the choose of the optimal *CI* is linked to the decisions of the hypothesis test, and it is preferable to choose a conservative test rather than a very liberal one (as there will be no false significances because its type I error is lower than the nominal one). The method for the *CIs* for the ratio $\theta$ is justified in a similar way.